\newcommand\aastexcls{2}
\newcommand\othercls{3}
\newcommand\papercls{\aastexcls}
\documentclass[tighten, times, twocolumn]{aastex62}

\usepackage{ifthen}
\usepackage{natbib}
\usepackage{amssymb, amsmath}
\usepackage{appendix}
\usepackage{etoolbox}
\usepackage[T1]{fontenc}
\usepackage{paralist}

\if\papercls \aastexcls
\hypersetup{citecolor=blue, 
            linkcolor=blue, 
            menucolor=blue, 
            urlcolor=blue}  
\else
\usepackage[
bookmarks=true,           
bookmarksnumbered=true,   
colorlinks=true,          
citecolor=blue,           
linkcolor=blue,           
menucolor=blue,           
urlcolor=blue,            
linkbordercolor={0 0 1},  
pdfborder={0 0 1},
frenchlinks=true]{hyperref}
\fi

\if\papercls \othercls

\else

\fi

\providecommand{\adsurl}[1]{\href{#1}{ADS}}

\makeatletter
\patchcmd{\NAT@citex}
  {\@citea\NAT@hyper@{%
     \NAT@nmfmt{\NAT@nm}%
     \hyper@natlinkbreak{\NAT@aysep\NAT@spacechar}{\@citeb\@extra@b@citeb}%
     \NAT@date}}
  {\@citea\NAT@nmfmt{\NAT@nm}%
   \NAT@aysep\NAT@spacechar\NAT@hyper@{\NAT@date}}{}{}

\patchcmd{\NAT@citex}
  {\@citea\NAT@hyper@{%
     \NAT@nmfmt{\NAT@nm}%
     \hyper@natlinkbreak{\NAT@spacechar\NAT@@open\if*#1*\else#1\NAT@spacechar\fi}%
       {\@citeb\@extra@b@citeb}%
     \NAT@date}}
  {\@citea\NAT@nmfmt{\NAT@nm}%
   \NAT@spacechar\NAT@@open\if*#1*\else#1\NAT@spacechar\fi\NAT@hyper@{\NAT@date}}
  {}{}
\makeatother

\makeatletter
\DeclareRobustCommand{\lowcase}[1]{\@lowcase#1\@nil}
\def\@lowcase#1\@nil{\if\relax#1\relax\else\MakeLowercase{#1}\fi}
\makeatother

\DeclareSymbolFont{UPM}{U}{eur}{m}{n}
\DeclareMathSymbol{\umu}{0}{UPM}{"16}
\let\oldumu=\umu
\renewcommand\umu{\ifmmode\oldumu\else\math{\oldumu}\fi}

\if\papercls \othercls

\else

\fi

\let\oldsim=\sim
\renewcommand\sim{\ifmmode\oldsim\else\math{\oldsim}\fi}
\let\oldpm=\pm
\renewcommand\pm{\ifmmode\oldpm\else\math{\oldpm}\fi}
\newcommand\by{\ifmmode\times\else\math{\times}\fi}

\newbox{\wdbox}
\renewcommand\c{\setbox\wdbox=\hbox{,}\hspace{\wd\wdbox}}
\renewcommand\i{\setbox\wdbox=\hbox{i}\hspace{\wd\wdbox}}

\newcount\timect
\newcount\hourct
\newcount\minct
\newcommand\now{\timect=\time \divide\timect by 60
         \hourct=\timect \multiply\hourct by 60
         \minct=\time \advance\minct by -\hourct
         \number\timect:\ifnum \minct < 10 0\fi\number\minct}

\catcode`@=11

\newcommand\comment[1]{}

\newcommand\commenton{\catcode`\%=14}

\renewcommand\math[1]{$#1$}
\newcommand\mathshifton{\catcode`\$=3}

\let\atab=&
\newcommand\atabon{\catcode`\&=4}

\let\oldmsp=\sp
\let\oldmsb=\sb
\def\sp#1{\ifmmode
           \oldmsp{#1}%
         \else\strut\raise.85ex\hbox{\scriptsize #1}\fi}
\def\sb#1{\ifmmode
           \oldmsb{#1}%
         \else\strut\raise-.54ex\hbox{\scriptsize #1}\fi}
\newbox\@sp
\newbox\@sb
\def\sbp#1#2{\ifmmode%
           \oldmsb{#1}\oldmsp{#2}%
         \else
           \setbox\@sb=\hbox{\sb{#1}}%
           \setbox\@sp=\hbox{\sp{#2}}%
           \rlap{\copy\@sb}\copy\@sp
           \ifdim \wd\@sb >\wd\@sp
             \hskip -\wd\@sp \hskip \wd\@sb
           \fi
        \fi}
\def\msp#1{\ifmmode
           \oldmsp{#1}
         \else \math{\oldmsp{#1}}\fi}
\def\msb#1{\ifmmode
           \oldmsb{#1}
         \else \math{\oldmsb{#1}}\fi}

\def\supon{\catcode`\^=7}

\def\subon{\catcode`\_=8}

\def\supsubon{\supon \subon}

\newcommand\actcharon{\catcode`\~=13}

\newcommand\paramon{\catcode`\#=6}

\comment{And now to turn us totally on and off...}

\newcommand\reservedcharson{ \commenton  \mathshifton  \atabon  \supsubon
                             \actcharon  \paramon}

\catcode`@=12
\reservedcharson

\newcommand\chisq{\ifmmode{\chi\sp{2}}\else\math{\chi\sp{2}}\fi}
\newcommand\redchisq{\ifmmode{ \chi\sp{2}\sb{\rm red}}
                    \else\math{\chi\sp{2}\sb{\rm red}}\fi}
\newcommand\Teq{\ifmmode{T\sb{\rm eq}}\else$T$\sb{eq}\fi}
\newcommand\mjup{\ifmmode{M\sb{\rm Jup}}\else$M$\sb{Jup}\fi}
\newcommand\rjup{\ifmmode{R\sb{\rm Jup}}\else$R$\sb{Jup}\fi}
\newcommand\msun{\ifmmode{M\sb{\odot}}\else$M\sb{\odot}$\fi}
\newcommand\rsun{\ifmmode{R\sb{\odot}}\else$R\sb{\odot}$\fi}
\newcommand\mearth{\ifmmode{M\sb{\oplus}}\else$M\sb{\oplus}$\fi}
\newcommand\rearth{\ifmmode{R\sb{\oplus}}\else$R\sb{\oplus}$\fi}

\begin{document}

\title{Ultra-Diffuse Galaxies at Ultraviolet Wavelengths}

\author{Pranjal RS}
\affiliation{Department of Physics, Indian Institute of Technology Bombay, Mumbai, 400076, India}
\author{Dennis Zaritsky} 
\affiliation{Steward Observatory \& Department of Astronomy, University of Arizona, Tucson, AZ 85719, USA}
\author{Richard Donnerstein}
\affiliation{Steward Observatory \& Department of Astronomy, University of Arizona, Tucson, AZ 85719, USA}
\author{Kristine Spekkens}
\affiliation{Department of Physics, Engineering Physics and Astronomy Queen's University Kingston, ON K7L 3N6, Canada}
\affiliation{Department of Physics and Space Science Royal Military College of Canada P.O. Box 17000, Station Forces Kingston, ON K7K 7B4, Canada}

\begin{abstract}
We measure NUV aperture magnitudes from {\sl GALEX} images for 258 ultra-diffuse galaxy (UDG) candidates drawn from the initial SMUDGes survey of $\sim 300$ square degrees surrounding, and including, the Coma galaxy cluster. For the vast majority, 242 of them, we present flux upper limits due either to a lack of significant flux in the aperture or confusion with other objects projected within the aperture. These limits often place interesting constraints on the UDG candidates, indicating that they are non-star forming or quiescent. In particular, we identify field, quiescent UDG candidates, which are a challenge for formation models and are therefore compelling prospects for spectroscopic follow-up and distance determinations. We present FUV and NUV magnitudes for 16 detected UDG candidates and compare those galaxies to the local population of galaxies on color-magnitude and specific star formation rate diagrams. The NUV detected UDG candidates form mostly an extension toward lower stellar masses of the star forming galaxy sequence and none of these lie within regions of high local galaxy density. UDG candidates span a range of properties, although almost all are consistent with being quiescent, low surface brightness galaxies, regardless of environment. 
\end{abstract}

\keywords{galaxies: fundamental parameters, structure}

\section{Introduction}

Physically large, and possibly massive, low surface brightness galaxies, recently coined as ultra-diffuse galaxies \citep[UDGs;][]{vdk15a}, are potentially unique testing grounds for theories of galaxy formation \citep[e.g.][]{yozin,amorisco,agertz,dic,rong,chan,carleton} and the nature of dark matter \citep[e.g.][]{bernal}. Low surface brightness galaxies, even physically large ones, have been known to exist for decades \citep[e.g.][]{sandage,impey,conselice}, but the novel and compelling development is that the largest among these appear to inhabit massive dark matter halos, such that their mass-to-light ratios are comparable to those of the most dark matter dominated dwarf spheroidal satellites of our Galaxy.  

Evidence supporting the large inferred masses comes from kinematic measurements of integrated starlight \citep{vdk16} and of globular clusters \citep{beasleya}, and from the size of their globular cluster populations \citep{vdk17}. There are caveats associated with the interpretation of both the kinematics, which are measured at small radii necessitating significant extrapolation to estimate total masses, and  the globular cluster counts, which adopt large completeness corrections and assume that the relationship between the number of globular clusters and total mass \citep{blakeslee,georgiev,harris,forbes,harris17,zar17b,forbes18} holds for UDGs. Even so, it is only the physically largest UDGs that appear to be this extreme in their total mass, while the physically smaller, more common among them appear to have lower masses that are consistent with those of dwarf galaxies \citep{amorisco1,sifon}. Therefore, much of the population of UDGs is likely to overlap what had been previously referred to as low surface brightness (LSB) galaxies \citep{disney,schombert,sch,sprayberry,dalcanton,conselice}. 

Large area photometric surveys are now uncovering LSBs/UDGs by the hundreds and thousands in a variety of environments \citep[eg.][]{yagi,remco,greco,smudges}. Such surveys are crucial if one wants to compile a large sample of the largest, most massive UDGs --- the objects that are truly unusual and most likely to place demanding constraints on both models of galaxy evolution and dark matter.  However, the lack of redshifts and other ancillary data for nearly all UDG candidates means that these objects are solely defined by surface brightness and angular size, leading to what is likely to be a heterogeneous population \citep{zar17,greco,ferre,lim}, thereby mitigating their value as probes of galaxy evolution and dark matter.

Measuring redshifts is currently the limiting factor in utilizing UDGs to advance our understanding of galaxy properties. Optical spectroscopy is expensive, requiring multiple hours per object on our largest ground-based telescopes \citep[eg.][]{vdk15a,kadowaki} and neutral hydrogen observations yield redshifts only for the small fraction of UDG candidates with gas  \citep{spekkens}. 
Guidance regarding the most promising candidates to target for the expensive spectroscopic follow-up would help mitigate the cost of the observations. 
Projected membership in a galaxy cluster or group has been one adopted approach, nearly guaranteeing that the adopted distance is the true distance \citep[eg.][]{kadowaki,alabi}. However, such galaxies have evolved within a  dynamical environment, complicating the interpretation of their evolution and structure.

Because of the common focus on UDGs in galaxy clusters, due to the question of distances and observational efficiency, those systems are better characterized than UDGs in low density environments. UDGs in clusters are almost exclusively red, consistent with being quiescent \citep{yagi,remco}.
This empirical finding has led to speculation that the evolution of these galaxies is guided by tidal effects, ram pressure gas loss, environmental ``strangulation", and basically all of the phenomena that are speculated to alter high brightness galaxies in clusters \citep[cf.][]{boselli}. 
In contrast, the small number of UDGs spectroscopically confirmed to lie in the field are optically bluer \citep{smudges} and sometimes have associated H{\small I} \citep{leisman,spekkens}. In fact, several UDG origin scenarios predict an absence of red UDGs in the field \citep[eg.][]{dic,chan}. Unfortunately, the field UDG population is poorly studied.

The use of existing ultraviolet imaging from the 
{\sl GALEX} \citep{GALEX} mission archive is an efficient way to measure the prevalence of recent star formation in UDGs across all environments. The use of UV flux to form a UV-optical color has the advantage that it is far more sensitive to recent star formation than are optical colors, while it has the disadvantage that it is in general much more challenging to measure. The {\sl GALEX} archive was already examined by \cite{greco} in relation to their catalog of low surface brightness galaxies. They find a high detection fraction, 76\%, among their optically blue sample, although the UDG selection criteria are different between their sample and SMUDGes. For example, \cite{greco} include systems of smaller angular extent and base their surface brightness criteria on a mean surface brightness within the effective radius rather than a central surface brightness.  
Here we investigate whether the set of UDG candidates released in the first SMUDGes catalog \citep{smudges} has NUV counterparts, what the properties of those systems are, and interpret the limits from those that are  not detected. In \S2 we describe our analysis of the {\sl GALEX} data and how we obtain our photometry. In \S3 we present our results, particularly the distribution of UDGs in color-magnitude space relative to ``normal" local galaxies and the spatial distribution of various UDG subclasses. 

\section{Data}
We begin with the UDG catalog from \cite{smudges} of 275 UDG candidates with half light radii, $r_h$, $\gtrsim$ 5.3$\arcsec$ that lie within roughly a 10$^\circ$ projected radius from the Coma galaxy cluster. One key difference between SMUDGes and other UDG catalogs is that the angular criterion corresponds to a physical effective radius, for those objects at the distance of the Coma cluster, of 2.5 kpc rather than the canonical 1.5 kpc. For example, in terms of angular size alone, we select objects whose minimum size is over twice that set by \cite{greco}. There is no physically motivated justification for either size criterion, but we choose to focus our effort on the candidates that are more likely to be physically larger. The projected virial radius of the Coma cluster is $\sim 1.7^\circ$ \citep{kubo} so the bulk of the survey volume is well outside of the cluster, with 67\% of the UDG candidates in the \cite{smudges} sample projected beyond the Coma cluster virial radius. 

From the \cite{smudges} catalog, we extract the coordinates, the half light radii, and the optical magnitudes. We choose to use the $r$-band primarily, although $g$ and $z$ are also measured in SMUDGes, for comparison to existing work and because $r$ provides a compromise between providing the longest wavelength baseline in combination with NUV and the highest signal-to-noise.

The NUV data come from the NASA public archive portal MAST\footnote{\href{https://archive.stsci.edu/}{archive.stsci.edu}} of the \textit{GALEX} mission \citep{GALEX}. \textit{GALEX} was a 50 cm diameter UV telescope capable of imaging the sky in FUV (1350-1750 {\AA}) and NUV (1750-2750 {\AA})  bands simultaneously. 
We exclude UDG candidates without available $r$ band or that lie outside available \textit{GALEX} images, and our final sample consists of 258 candidates. As discussed below, detections were sufficiently uncommon in the NUV band that we do not present results for the FUV band for the full sample. The image database is not of uniform depth and so detection thresholds vary across the sample.

To determine whether we detect each individual candidate in the available data, 
we used the Astropy \citep{astropy} affiliated package Photutils\footnote{\href{http://doi.org/10.5281/zenodo.1039309}{doi.org/10.5281/zenodo.1039309}} to perform photometry on the NUV data at each candidate's location. We define a circular aperture of radius 2$r_{e}$, where $r_{e}$ is the optical half light radius of the target, and measure the flux in the NUV intensity map from which we subtract the background flux.
We determine the background flux by measuring the flux within the same circular aperture in the {\sl GALEX}-provided sky background image. We convert from counts per second (CPS) to an NUV magnitude in the AB system \citep{oke1,oke2} using the equation 
\begin{equation}
    m_{AB} = -2.5\times \log_{10}(CPS) + 20.08,
\end{equation}
\citep{GALEX_mag}.

To check that the background image is uncontaminated by the UDG candidate itself 
and to account for the effect of variations in the background map, we recompute $m_{NUV}$ after displacing the background aperture, starting from a displacement of 2$r_h$ and increasing to 10$r_h$ in $\alpha$ and $\delta$. Although the change in magnitude tends to increase with the displacement, the absolute value of the difference is $<$ 0.1 mag for the vast majority (ranging from 240 out of 258 for the smallest displacements to 200 out of 258 for the largest).

The point spread function (PSF) for the GALEX NUV images is 6.0$\arcsec$ FWHM \citep{GALEX}, which can lead to confusion among sources and strong contamination in some cases. In our first step to mitigate this problem, we compute the ratio of the flux obtained within an aperture of radius 2$r_h$ to that obtained within an aperture of radius $r_h$. We reject candidates where this ratio is $>$ 4. We visually confirmed that this criterion identifies strongly contaminated objects. Fifty-three candidates are classified as strongly contaminated and not considered further. Other candidates where this ratio is $<$ 4 may nevertheless suffer more subtle contamination and we discuss that problem further below.

Among the remaining sources, we expect our uncertainties to be dominated by the uncertainty in the background determination. To estimate the mean background uncertainty, we measure the scatter in fluxes obtained from visually-inspected uncontaminated sky regions. We identify $40$ blank apertures (of equal size) and measure the enclosed flux. We then compute the standard deviation, $\sigma_{sky}$, by fitting a Gaussian to the resulting distribution of measured fluxes among the apertures. We define the mean background uncertainty per pixel to be $\sigma_{sky}\sqrt{N_{sky}}$, where $N_{sky}$ is the number of pixels in the sky aperture. The uncertainty in the mean background level within a target aperture is then $\sigma_{sky}\sqrt{N_{sky}}/\sqrt{N_{target}}$, where $N_{target}$ is the number of pixels in the target aperture. If there is a large scale gradient to the background, then our method for estimating the uncertainties will overestimate the background uncertainty locally. However, we cannot be sure that the background gradient is real or systematic and our method captures this uncertainty.

\begin{deluxetable}{lrr}
\tabletypesize{\scriptsize}
\tablewidth{0pt}
\tablecaption{Candidate Sorting
\label{tab:cuts}}
\tablehead{
\colhead{Criterion} &
\colhead{Number Rejected}&
\colhead{Number Remaining}\\
}
\startdata
Lacking NUV or r photometry&17&258\\
Contaminated (flux ratio)&53&205\\
SNR $<$ 2&168&37\\
Contaminated (visually determined)&21&16\\
\enddata
\end{deluxetable}

To ascertain our confidence in a source detection, we define the signal to noise ratio (\textit{SNR}) for each candidate UDG detection,  ignoring the uncertainty contribution from the source itself and any faint contaminating objects, as 
\begin{equation}
\textit{SNR} = \frac{\text{candidate flux}}{\text{background uncertainty}}.
\end{equation}
We consider sources with \textit{SNR}$<$2 to be non-detections and
visually inspect all others. While visually inspecting the candidates with SNR$\ge$2, we consider how well centered the object is, the extent of the object, and the possible contaminating effects of the neighbors on the photometry.
If we conclude that the detection cannot be confidently assigned to the UDG candidate, we label that as a non-detection. 
For all sources that are classified as non-detections, we either accept the measured flux as an upper limit on the flux or, in cases, where the measured flux is negative, we assign a value of twice the background uncertainty to be the flux upper limit. In cases where we suspect contamination, the measured flux is a conservative upper limit because we have not attempted to correct for the contaminating object.  

\begin{figure*}[!htbp]
 \begin{center}
\includegraphics[width = 0.20 \textwidth]{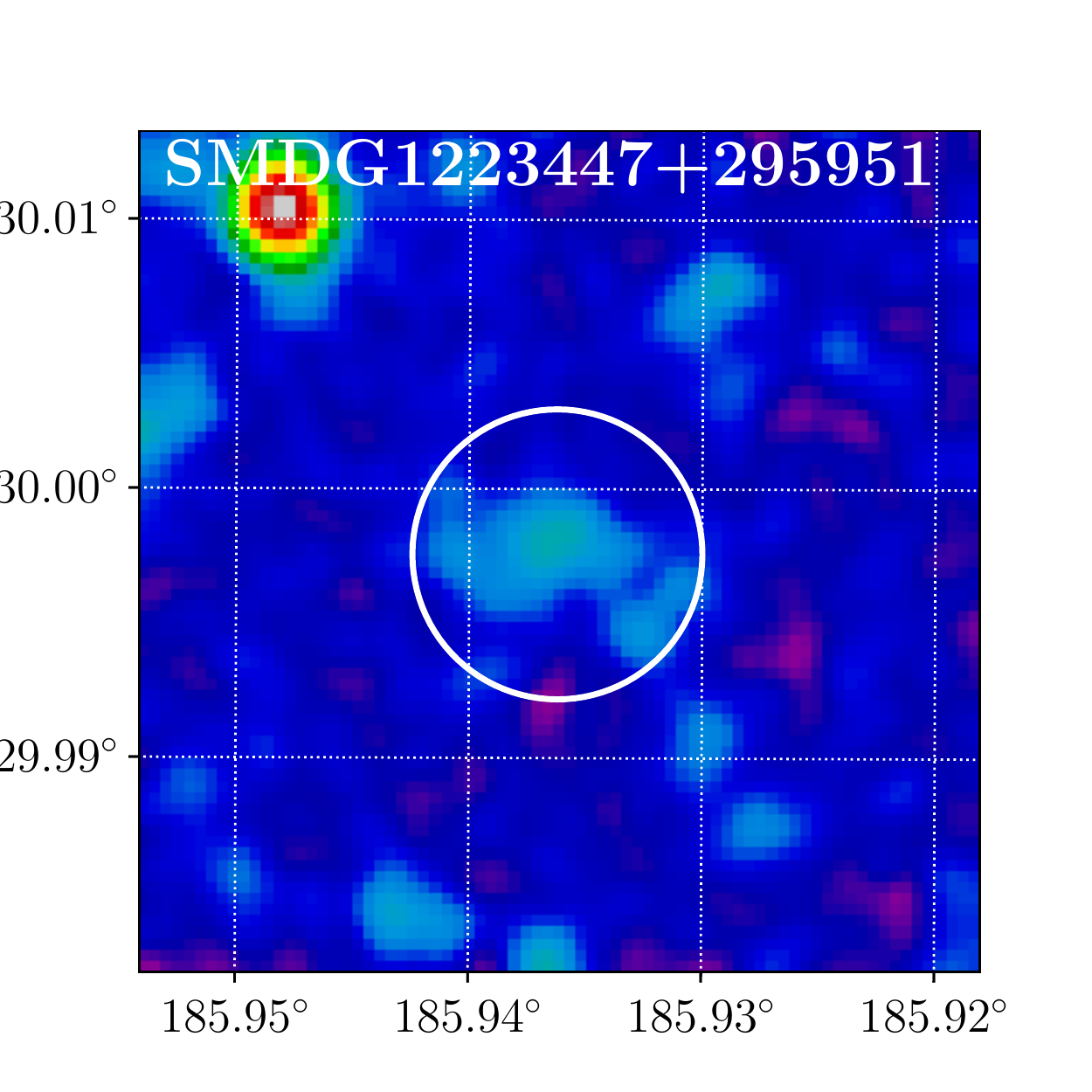}
\includegraphics[width = 0.20 \textwidth]{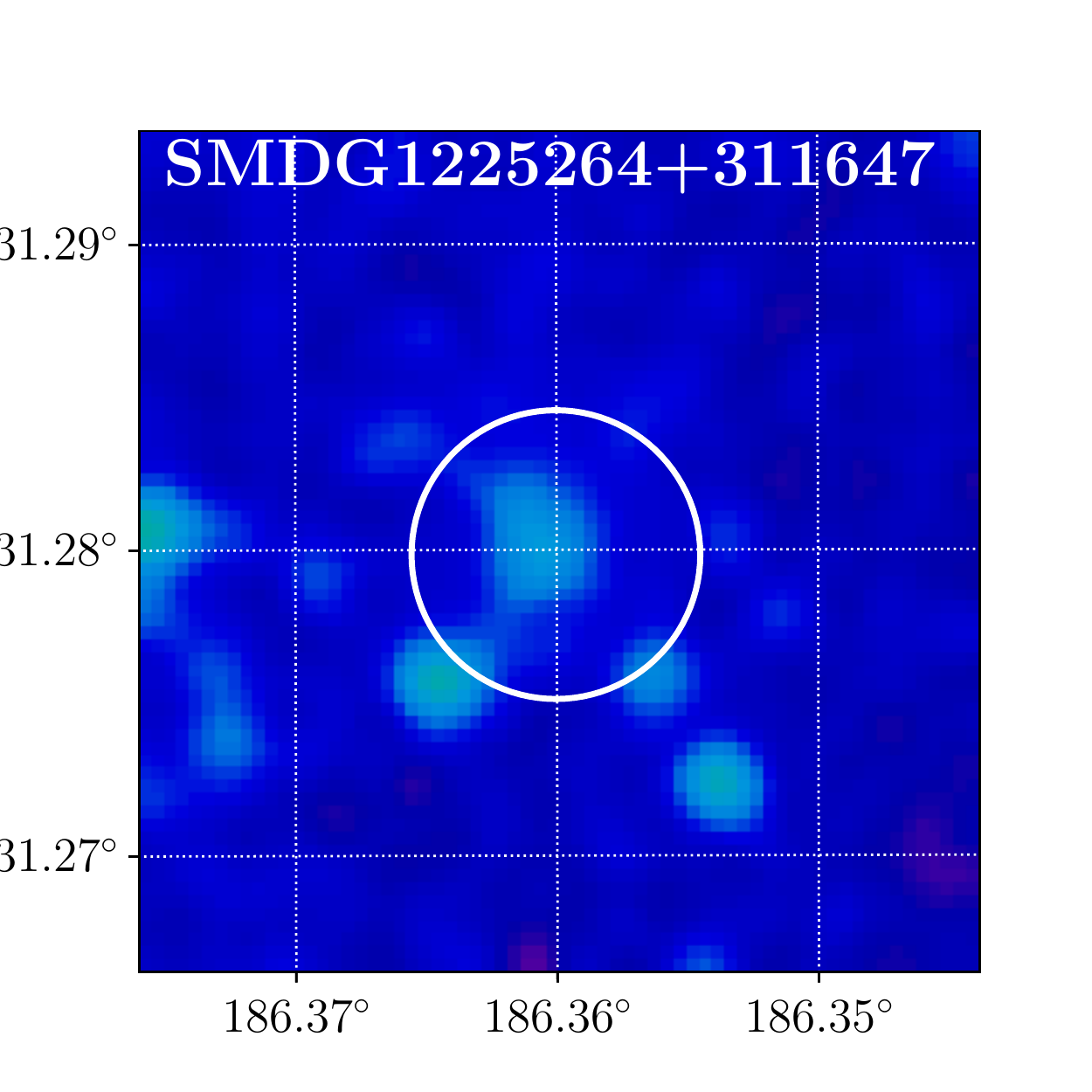}
 \includegraphics[width = 0.20 \textwidth]{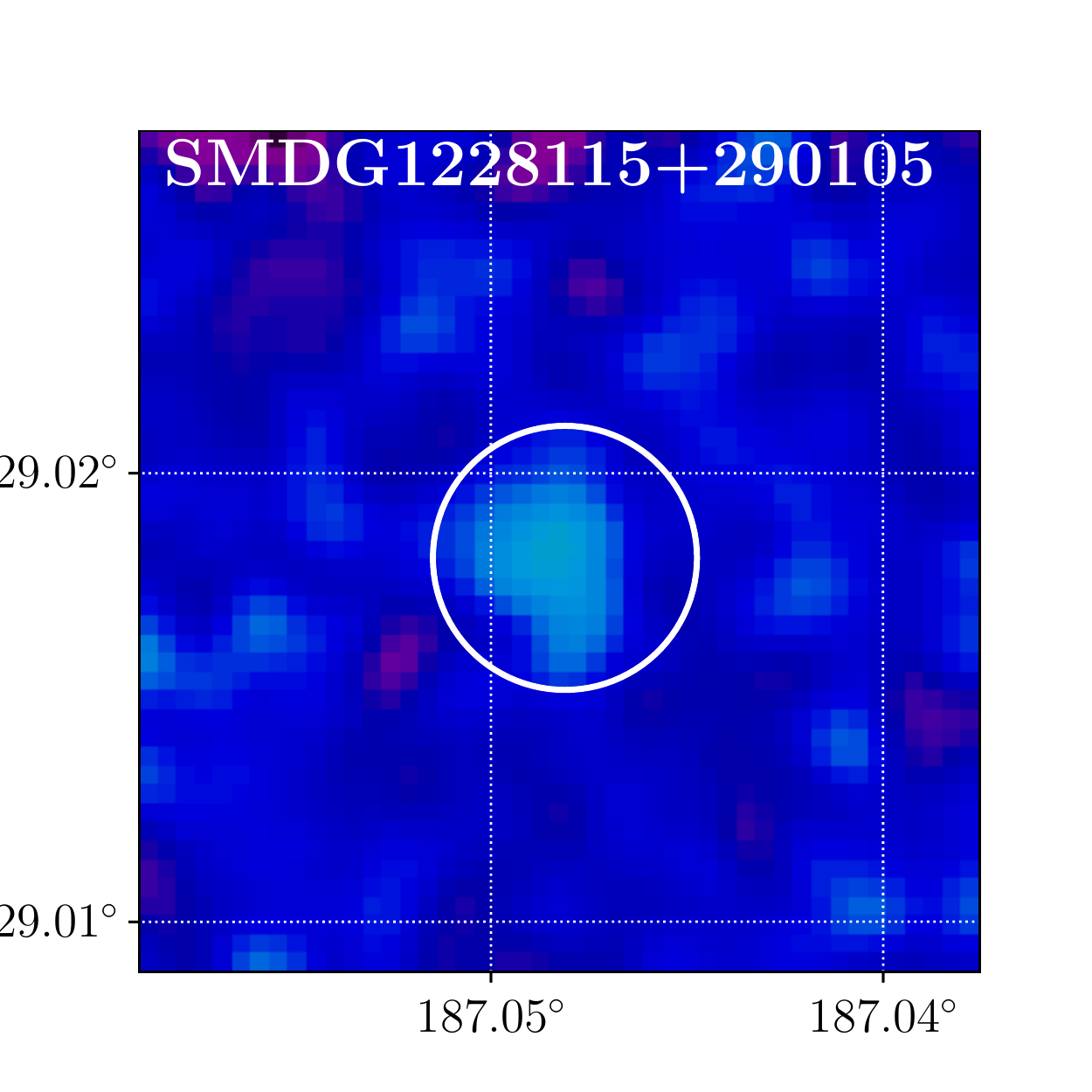}
 \includegraphics[width = 0.20 \textwidth]{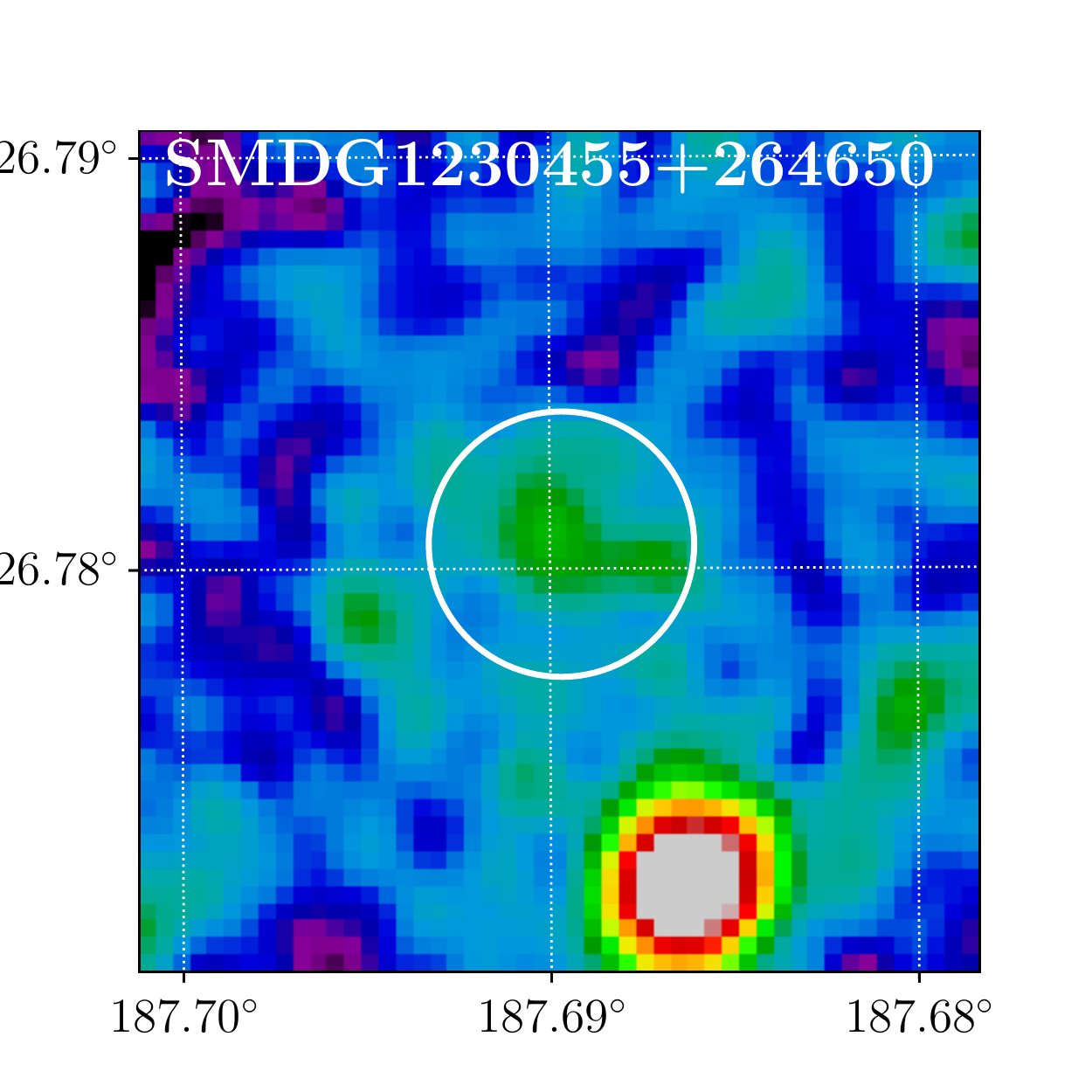}
 \includegraphics[width = 0.20 \textwidth]{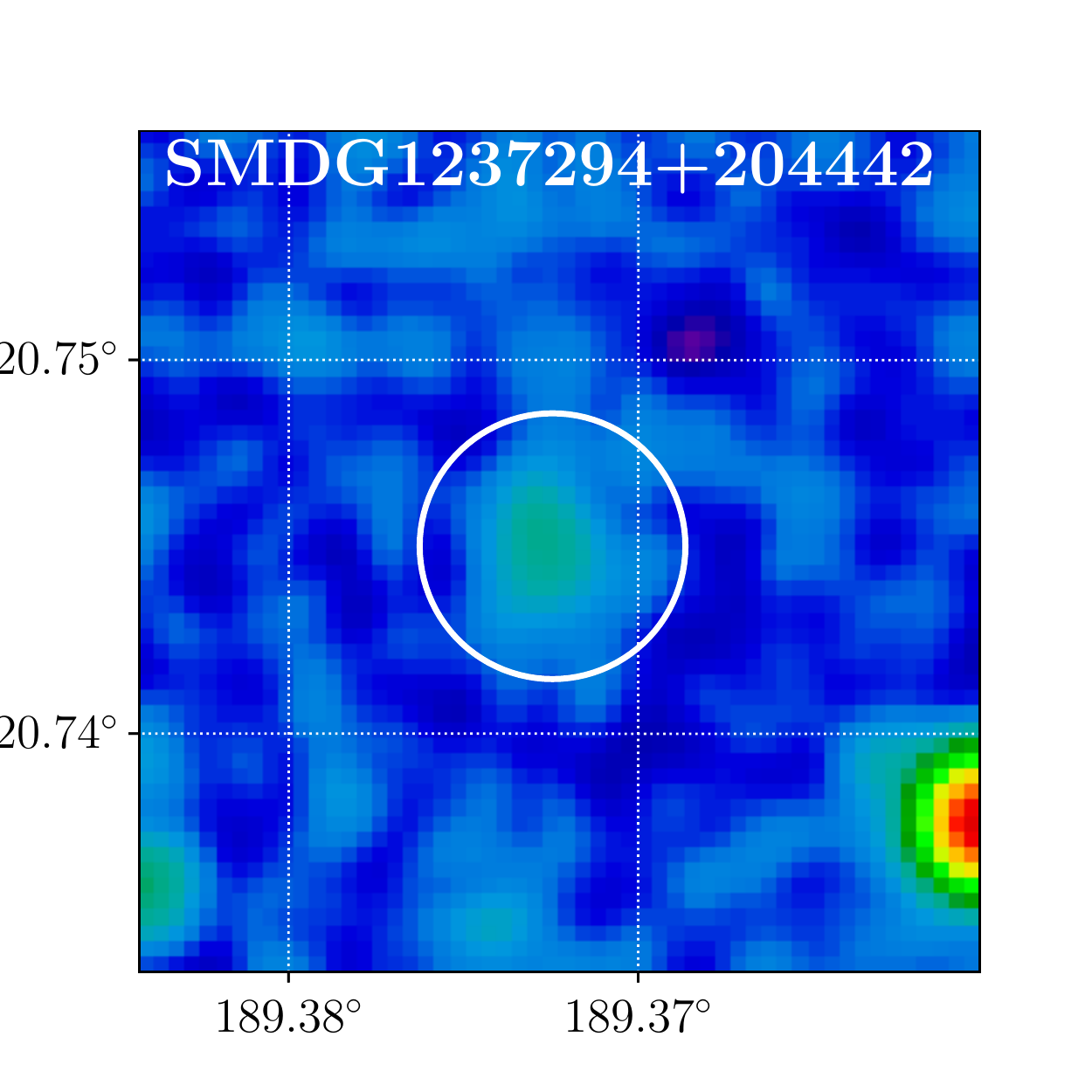}
 \includegraphics[width = 0.20 \textwidth]{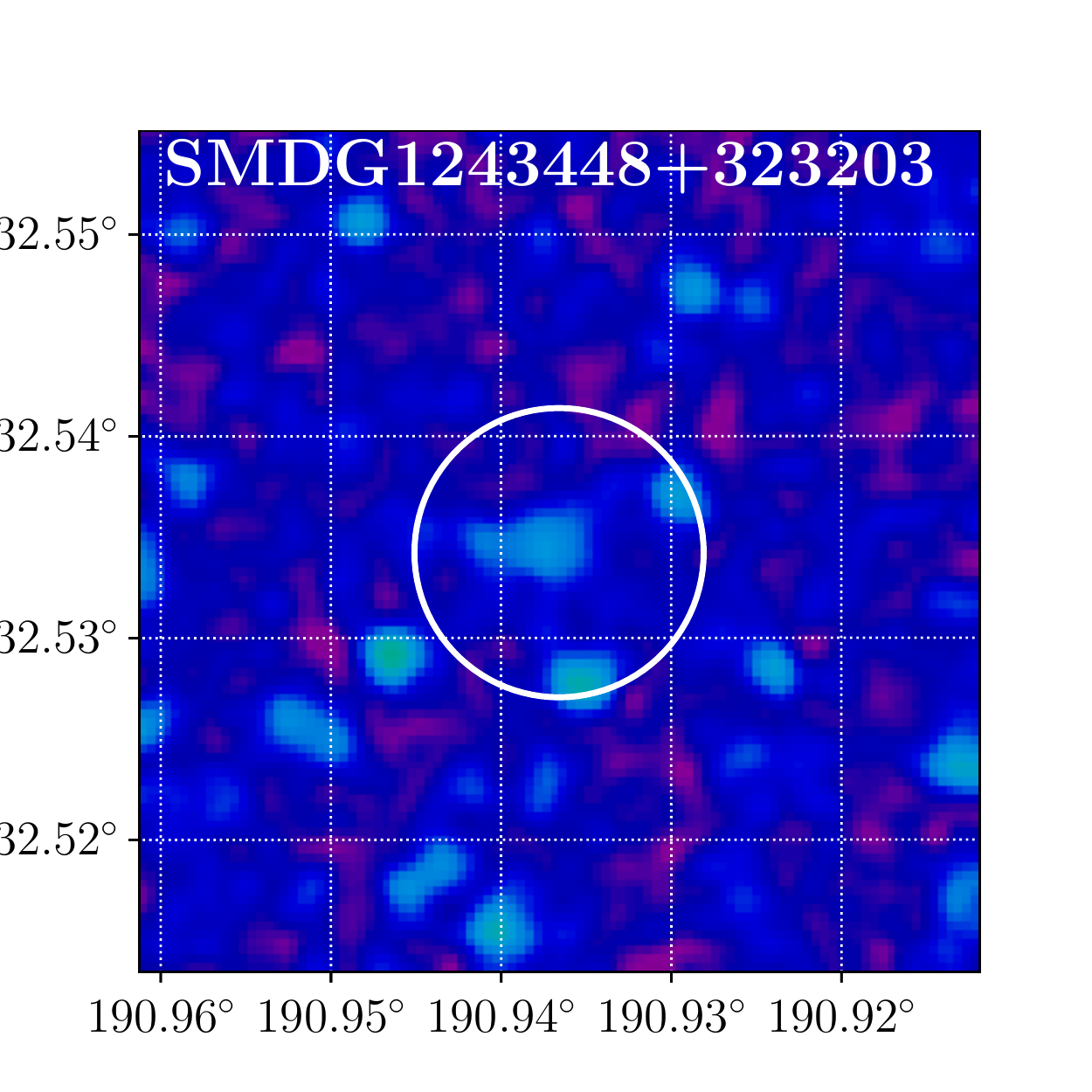}
 \includegraphics[width = 0.20 \textwidth]{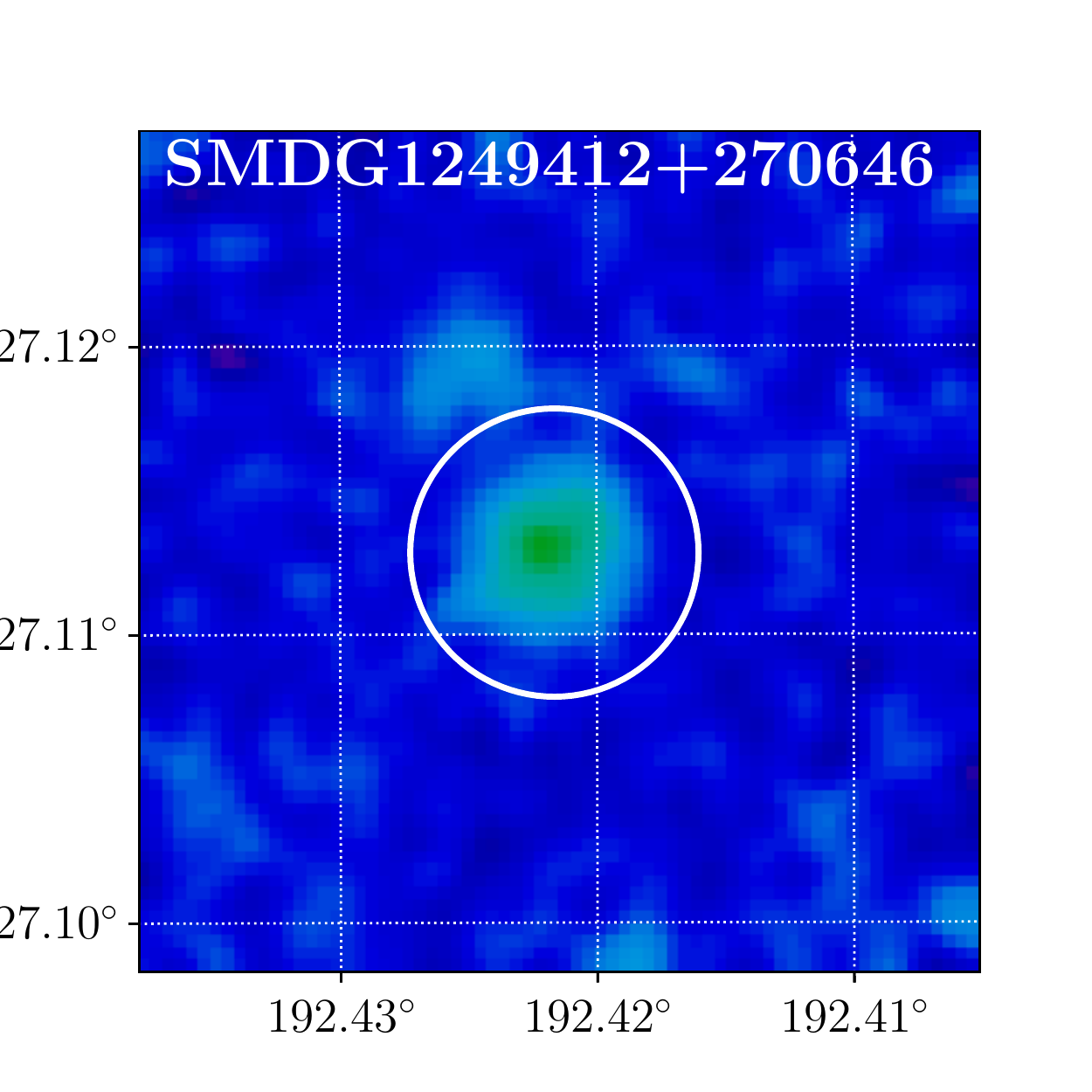}
 \includegraphics[width = 0.20 \textwidth]{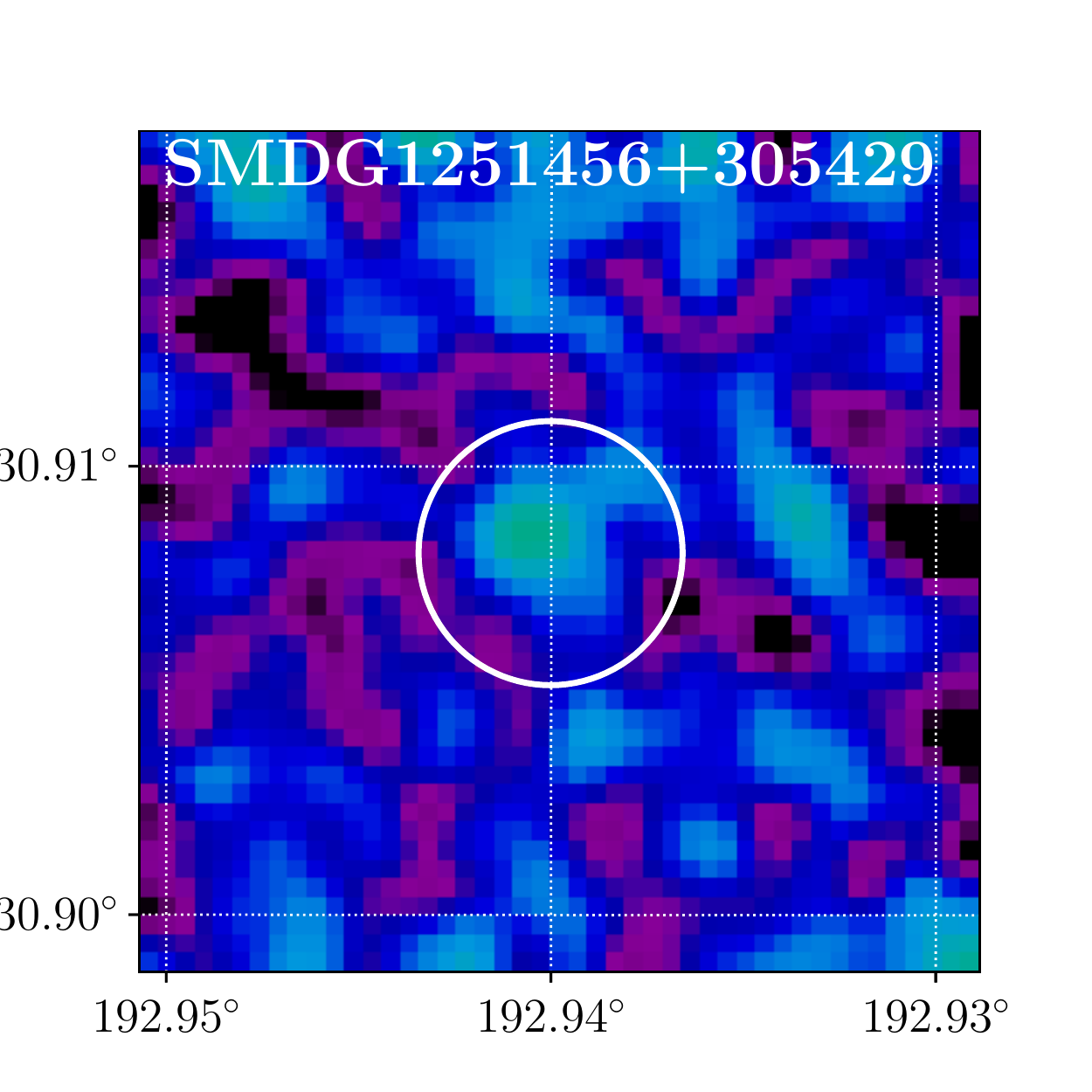}
 \includegraphics[width = 0.20 \textwidth]{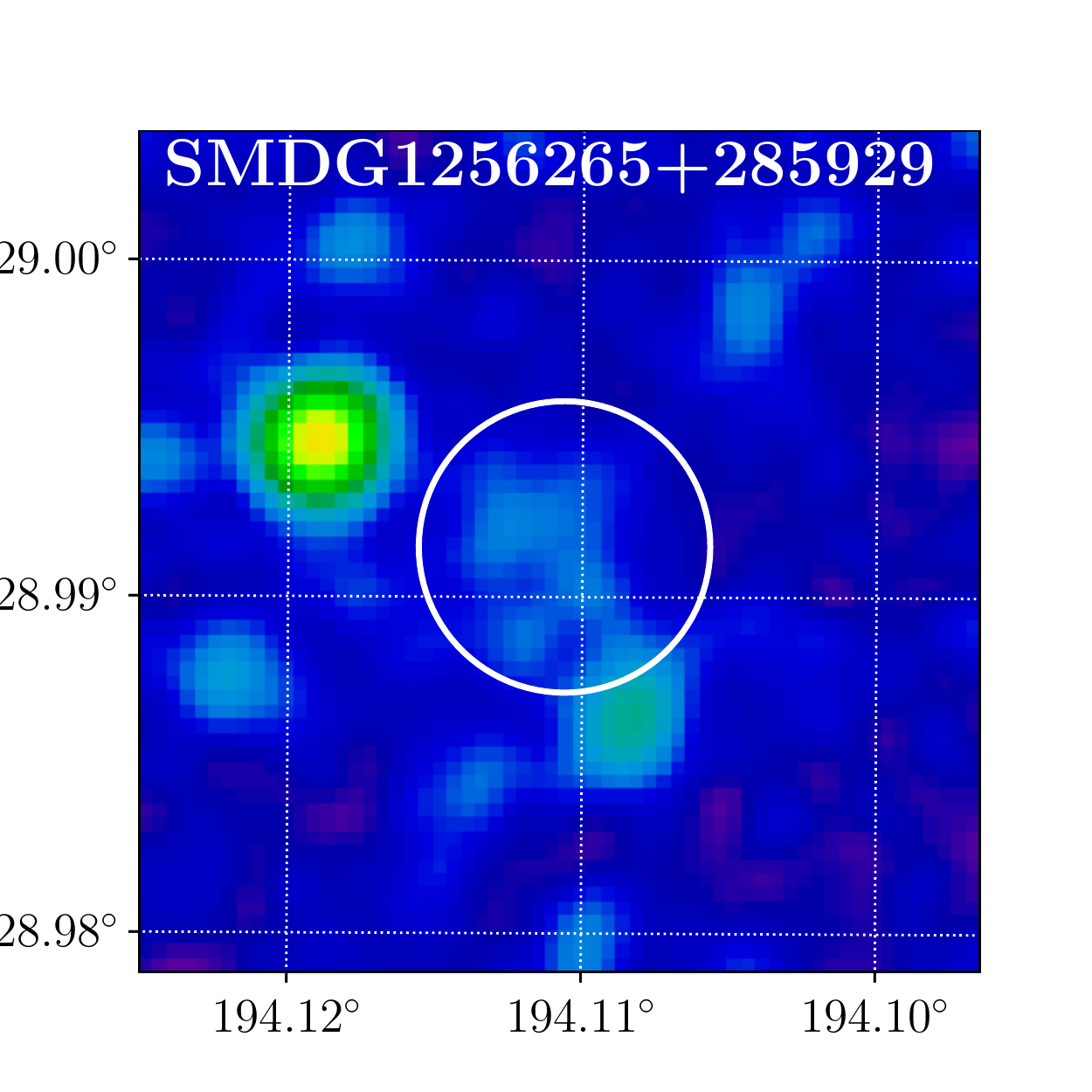}
 \includegraphics[width = 0.20 \textwidth]{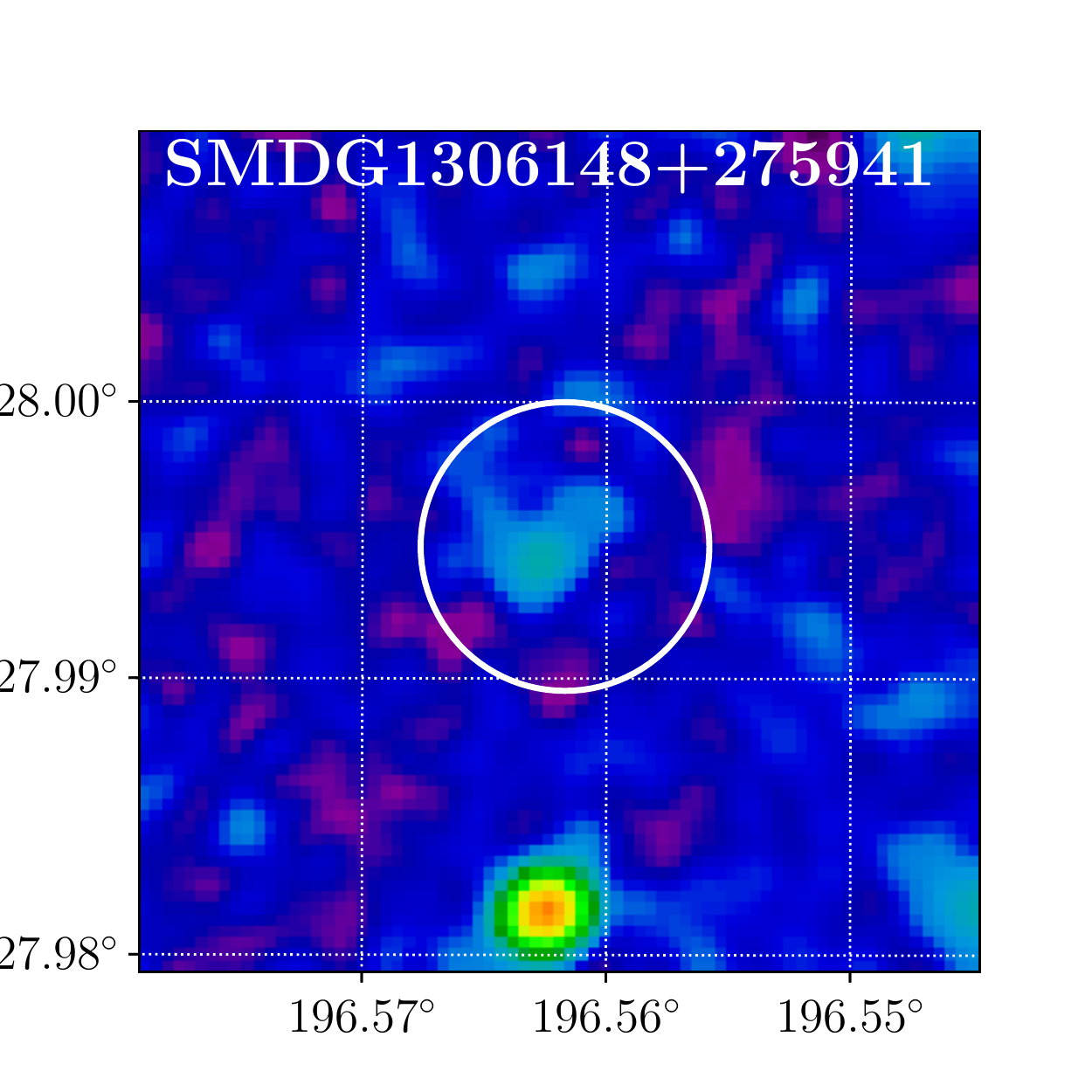}
 \includegraphics[width = 0.20 \textwidth]{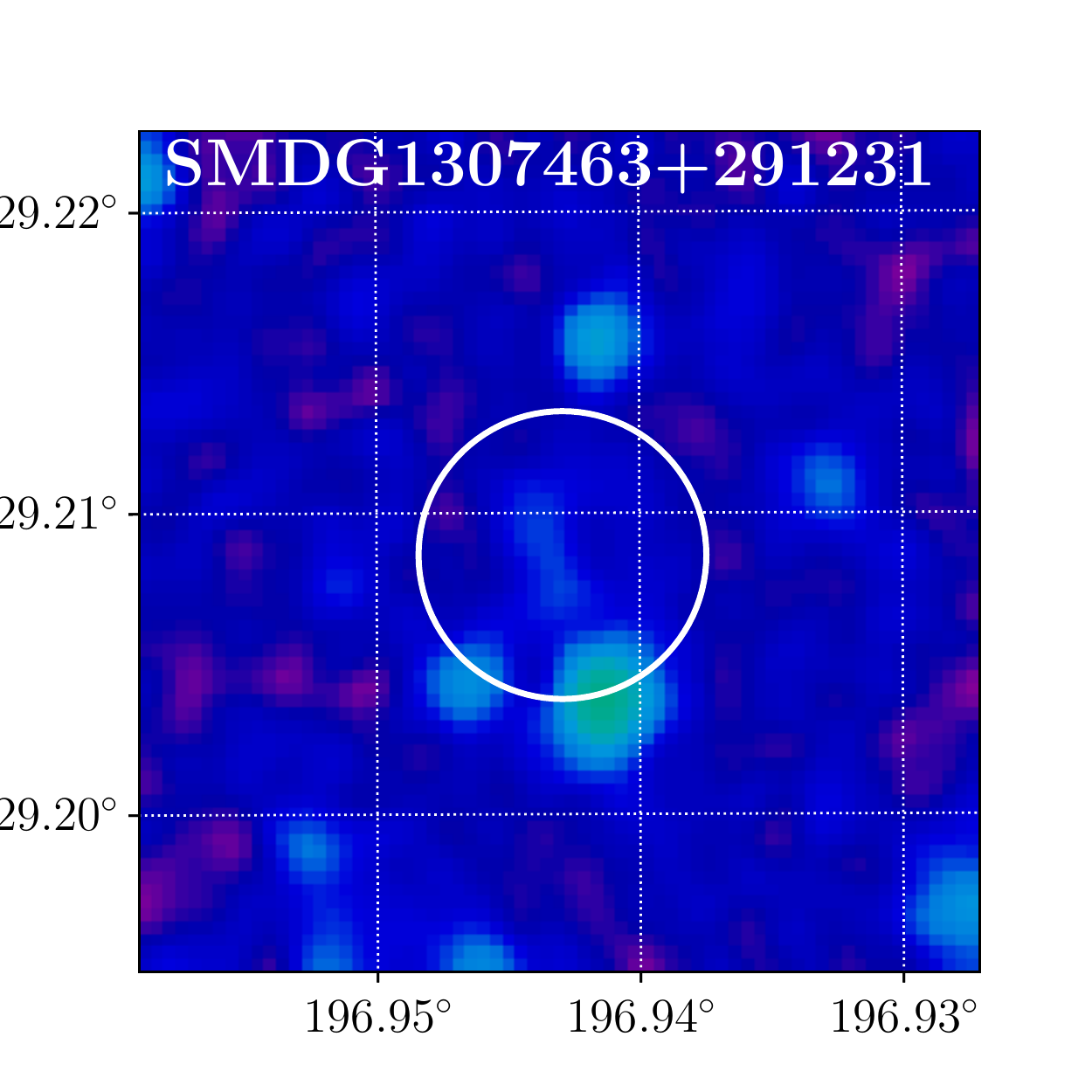}
 \includegraphics[width = 0.20 \textwidth]{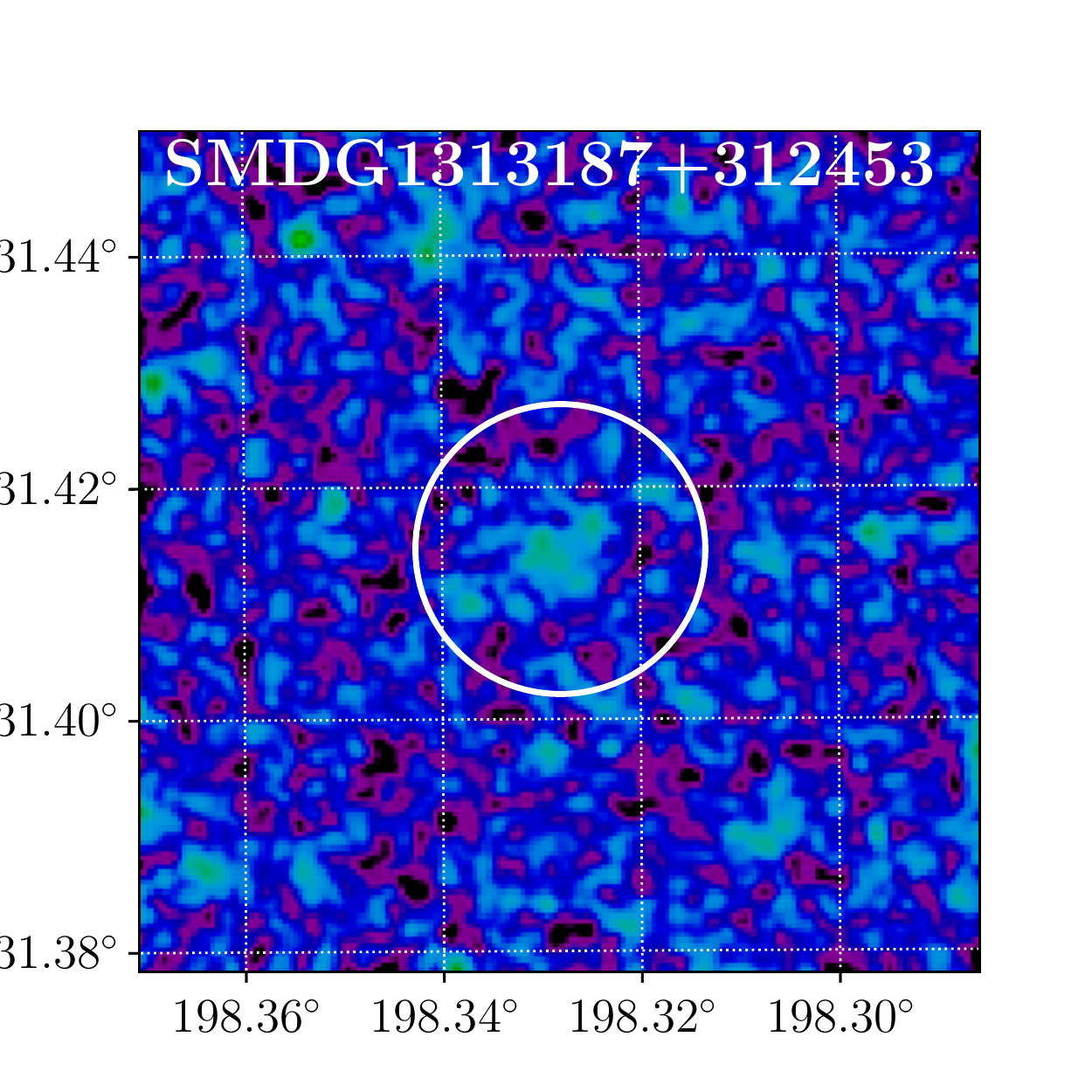}
 \includegraphics[width = 0.20 \textwidth]{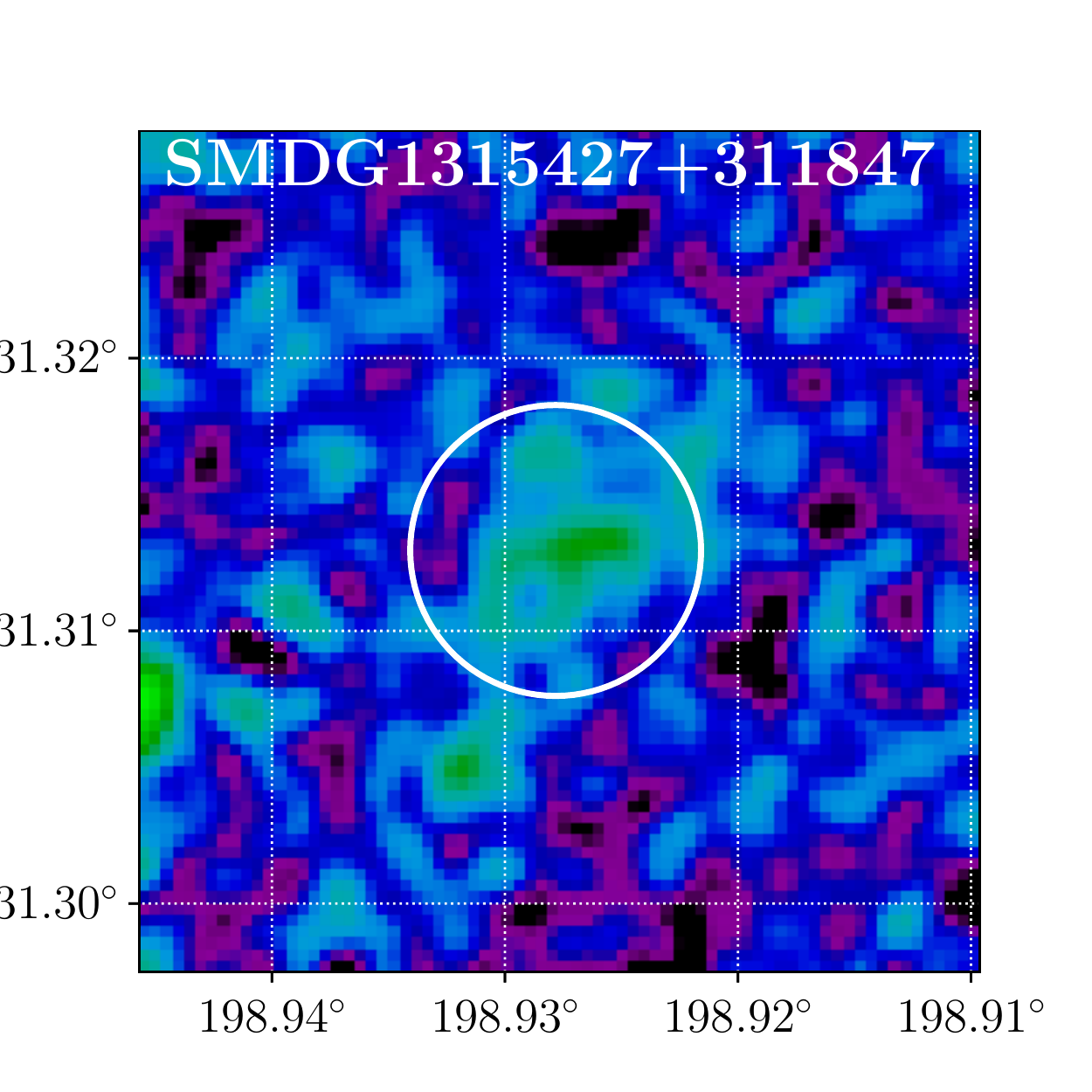}
 \includegraphics[width = 0.20 \textwidth]{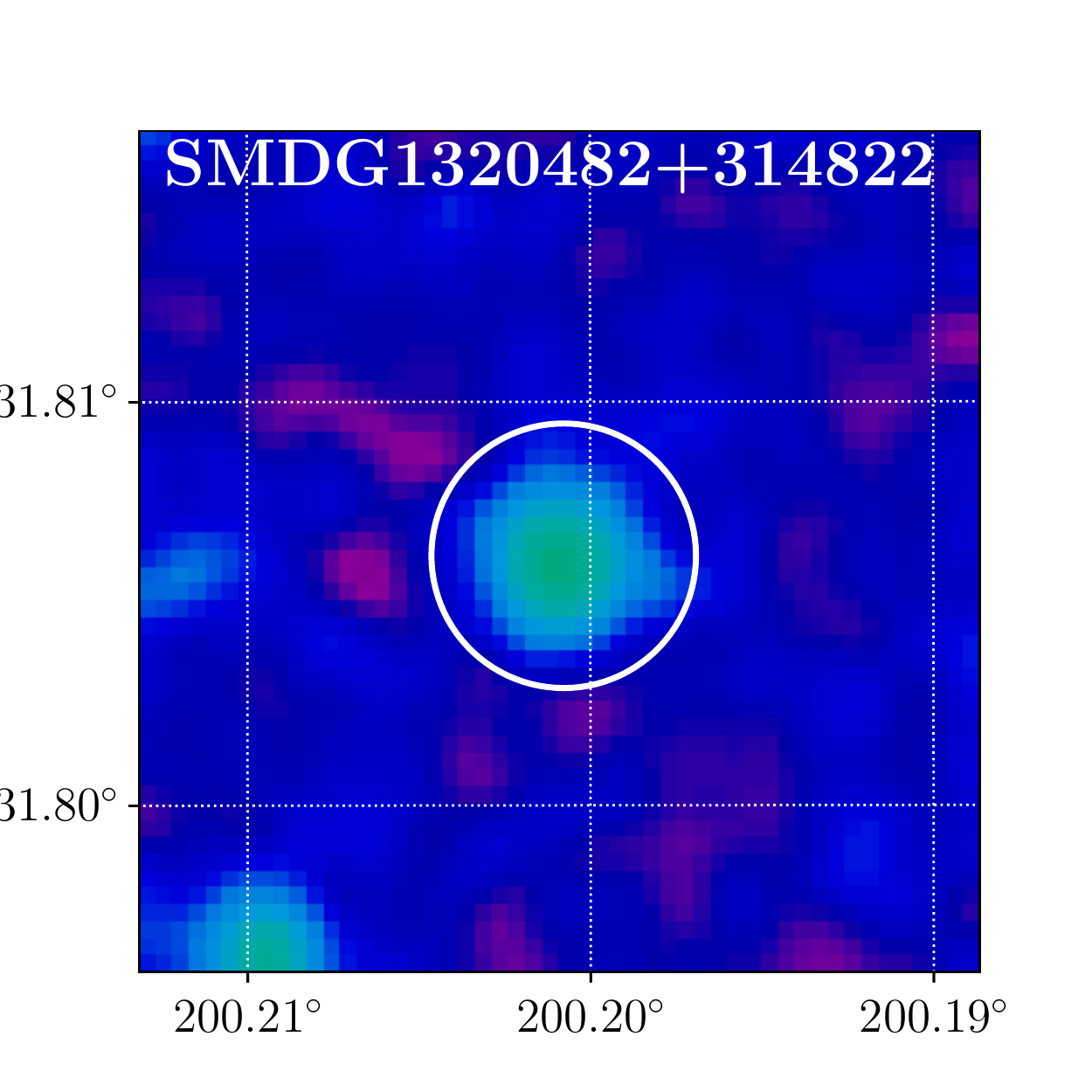}
 \includegraphics[width = 0.20 \textwidth]{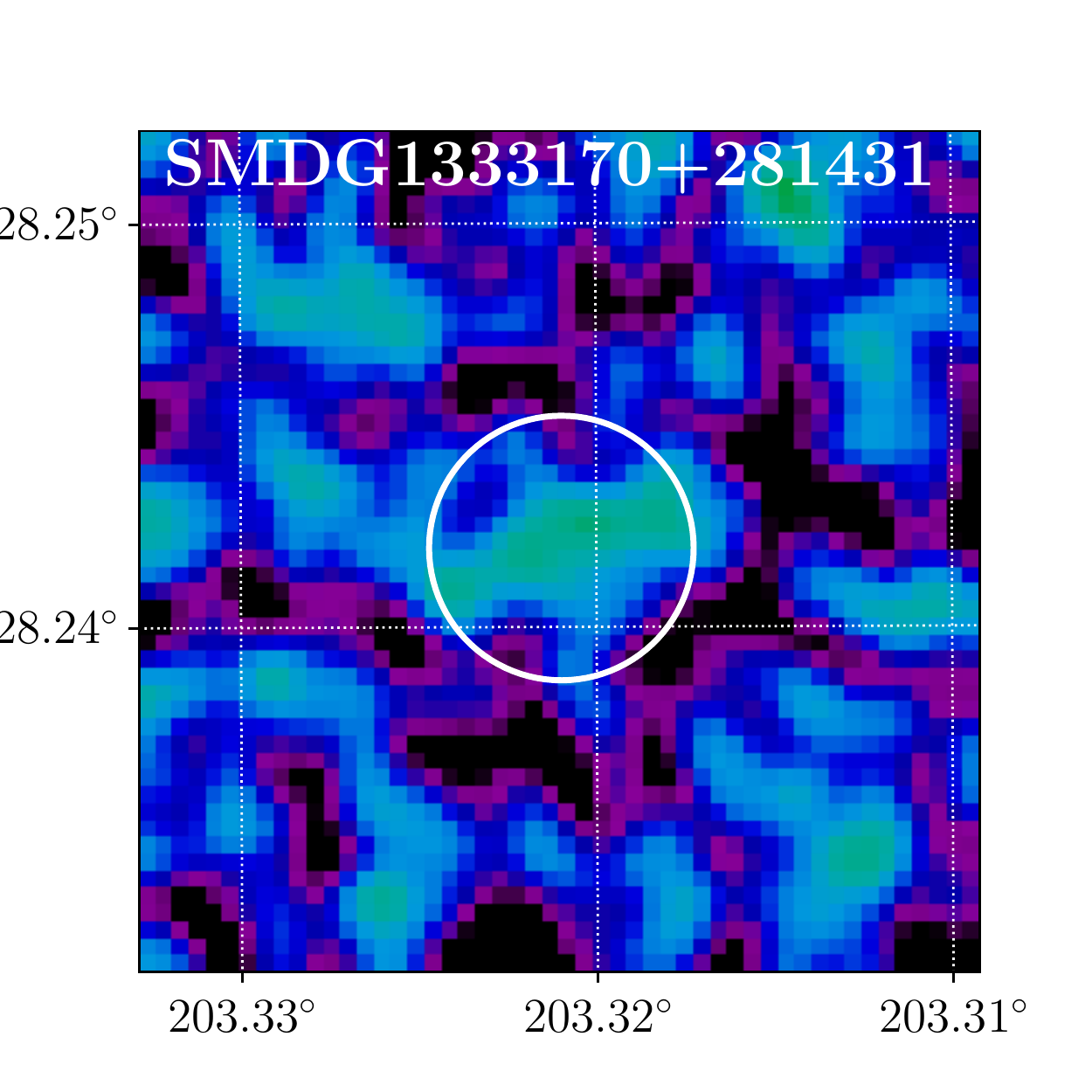}
 \includegraphics[width = 0.20 \textwidth]{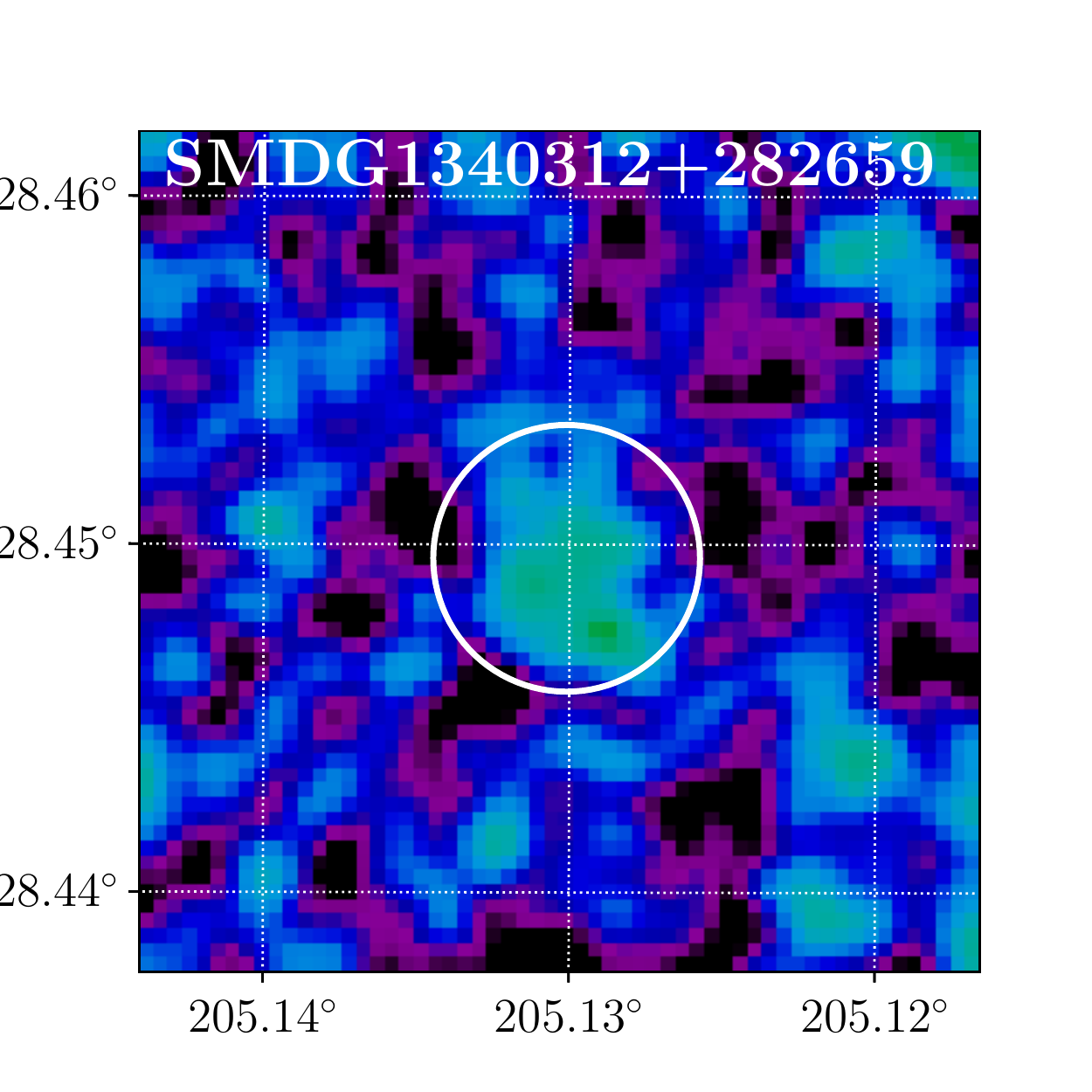}
\caption{Images of the 16 candidate UDGs for which we measure a significant NUV flux that is visually related to the optical detection. The angular scale of the images vary, but coordinates are given. The white circles are centered on the optical location of the UDG candidate and have a radius of 2$r_h$. Candidates are presented in order of increasing right ascension from left to right and downward.}
\label{fig:mosaic}
\end{center}
 \end{figure*}

After rejecting 53 candidates on the basis of our initial contamination criterion and 17 candidates without either NUV images or $r-$band data, we have 16 detections and 189 non-detections (the effect of different criteria on the final set of detected candidates is summarized in Table \ref{tab:cuts}).
The NUV images for the 16 detections are presented in Figure \ref{fig:mosaic}. For those detected in NUV, we also measure the FUV magnitude using the same procedure.
In Table \ref{tab:data} we present the extinction corrected values of 
the NUV and FUV magnitudes for those same UDG candidates and $r-$band magnitudes using published reddening maps \citep{Schlafy}
and the following selective extinction relations 
\begin{align}
    &A_{NUV}/E(B-V) = 8.2, \quad A_{FUV}/E(B-V) = 8.24, \notag \\
    &\mathrm{and}\quad 
    A_{r}/E(B-V) = 2.28,
\end{align}
as also adopted by \cite{Colour_extn}. We present the lower limits on $m_{FUV}$ for the objects that are not visually discernable in the FUV band. Table 3
uses a similar format and contains the extinction corrected $m_{NUV}$ limits for the non-detections.

Some of the detections could be chance coincidences with other UV-emitting objects. To gauge the magnitude of this effect, we placed random apertures within the same set of {\sl GALEX} images and found that $\sim$ 15\% of the time the detected flux satisfies our initial SNR $>$ 2 criterion. After visual examination, only $\sim$ 20\% of those are sufficiently well-centered and uncontaminated. In combination, these results suggest that 3\% of our UDG candidates (8 out of 258) could have false superpositions. Accounting for the fraction of contaminated sources, we conclude that 6 of our 16 detections might be attributable to superpositions.  

In contrast, some of the non-detections could be chance coincidences of emitting sources with strong, negative noise fluctuations. However, given our few  detections, we do not expect there to be a significant population of such objects. Assuming that the positive superpositions discussed above are matched by negative ones, which is overly conservative because  real sources contribute to positive fluctuations but not to negative ones, we  cap the fraction of such events at $\sim$ 15\%. Given 10 to 16 detected sources, we  expect to have ``lost" one to two sources that we would have otherwise detected.

\section{Results}

\begin{deluxetable}{crcrrr}
\tabletypesize{\scriptsize}
\tablewidth{0pt}
\tablecaption{NUV-Detected UDG Candidates Photometry
\label{tab:data}}
\tablehead{
\colhead{Object Name} &
\colhead{$r_h$}& 
\colhead{$m_r$}&
\colhead{$m_{NUV}$} &
\colhead{$m_{FUV}$}&
\colhead{$E(B-V)$}\\
&
\colhead{[arcsec]}}
\startdata
SMDG1223447+295951 & 9.7 & 18.39 & 21.11 \pm{0.26} & $>$ 23.54 \pm{0.61} & 0.02 \\
SMDG1225264+311647 & 8.5 & 17.62 & 21.68 \pm{0.38} & $>$ 22.07 \pm{0.14} & 0.02 \\
SMDG1228115+290105 & 5.3 & 20.04 & 22.56 \pm{0.55} & 22.98 \pm{0.20}& 0.02 \\
SMDG1230455+264650 & 5.8 & 19.02 & 20.64 \pm{0.10} & $>$ 22.31 \pm{0.12} & 0.02 \\
SMDG1237294+204442 & 6.4 & 19.56 & 21.64 \pm{0.30} & $>$ 24.20 \pm{0.80} & 0.03 \\
SMDG1243448+323203 & 12.9 & 17.88 & 21.11 \pm{0.33} & $>$ 23.85 \pm{1.05} & 0.01 \\
SMDG1249412+270646 & 9.0 & 18.98 & 20.89 \pm{0.19} & 22.86 \pm{0.29} & 0.01 \\
SMDG1251456+305429 & 5.3 & 21.06 & 22.55 \pm{0.51} & $>$ 25.59 \pm{2.13} & 0.01 \\
SMDG1256265+285929 & 7.8 & 19.11 & 21.89 \pm{0.40} & $>$ 23.41 \pm{0.41} & 0.01 \\
SMDG1306148+275941 & 9.4 & 19.14 & 21.81 \pm{0.45} & 23.32 \pm{0.45}& 0.01 \\
SMDG1307463+291231 & 8.6 & 19.09 & 22.14 \pm{0.55} & $>$ 24.70 \pm{1.45} & 0.01 \\
SMDG1313187+312453 & 22.5 & 16.47 & 20.21 \pm{0.25} & $>$ 21.96 \pm{0.31} & 0.01 \\
SMDG1315427+311847 & 9.6 & 18.51 & 20.18 \pm{0.10} & $>$ 21.34 \pm{0.07} & 0.01 \\
SMDG1320482+314822 & 5.9 & 19.58 & 21.56 \pm{0.23} & 22.86 \pm{0.19} & 0.01 \\
SMDG1333170+281431 & 5.9 & 19.51 & 21.27 \pm{0.18} & 22.73 \pm{0.17} & 0.01 \\
SMDG1340312+282659 & 6.9 & 19.86 & 21.22 \pm{0.19} & 23.92 \pm{0.59} & 0.01 \\
\enddata
\end{deluxetable}

\begin{deluxetable}{crcrrr}
\tabletypesize{\scriptsize}
\tablewidth{0pt}
\tablecaption{NUV-Undetected UDG Candidates Photometry\tablenotemark{a}
\label{tab:nondetections}}
\tablehead{
\colhead{Object Name} &
\colhead{$r_h$}& 
\colhead{$m_{NUV} >$}&
\colhead{$m_r$} &
\colhead{$E(B-V)$}&
\colhead{$SNR <$}\\
&
\colhead{[arcsec]}}
\startdata
SMDG1212080+281630& 5.6&  22.7&  20.9& 0.02&   1.7\\
SMDG1212085+290348& 6.1&  25.6&  19.7& 0.02&   0.1\\
SMDG1212454+273506& 5.4&  23.6&  20.7& 0.02&   0.8\\
SMDG1213061+294551&10.6&  20.9&  18.0& 0.02&   4.6\\
SMDG1213235+264641& 6.4&  23.3&  20.6& 0.02&   0.9\\
SMDG1213512+282109& 5.5&  22.4&  19.8& 0.01&   2.2\\
SMDG1214010+293203& 6.3&  22.8&  20.8& 0.02&   1.3\\
SMDG1214279+294033& 6.1&  22.4&  20.6& 0.02&  $-$0.3\\
SMDG1214418+274954& 8.4&  22.0&  18.7& 0.02&   0.0\\
SMDG1214429+291508&11.3&  22.7&  18.6& 0.02&   0.9\\
\enddata
\tablenote{Only the first 10 entries are presented here for reference. The complete table is available electronically.}
\end{deluxetable}

NUV detections from candidate UDGs drawn from the SMUDGes catalog in {\sl GALEX} images are rare. We identify only 16 out of 258 (6\%), or only 10 (4\%) if we account for the possibility of source superposition. Of course, there are a variety of possible reasons for why the remaining systems have low UV fluxes, but even the limits convey information.

\subsection{Distribution in Color-Magnitude Space}

\begin{figure}[!htbp]
\begin{center}
\includegraphics[width = 0.48 \textwidth]{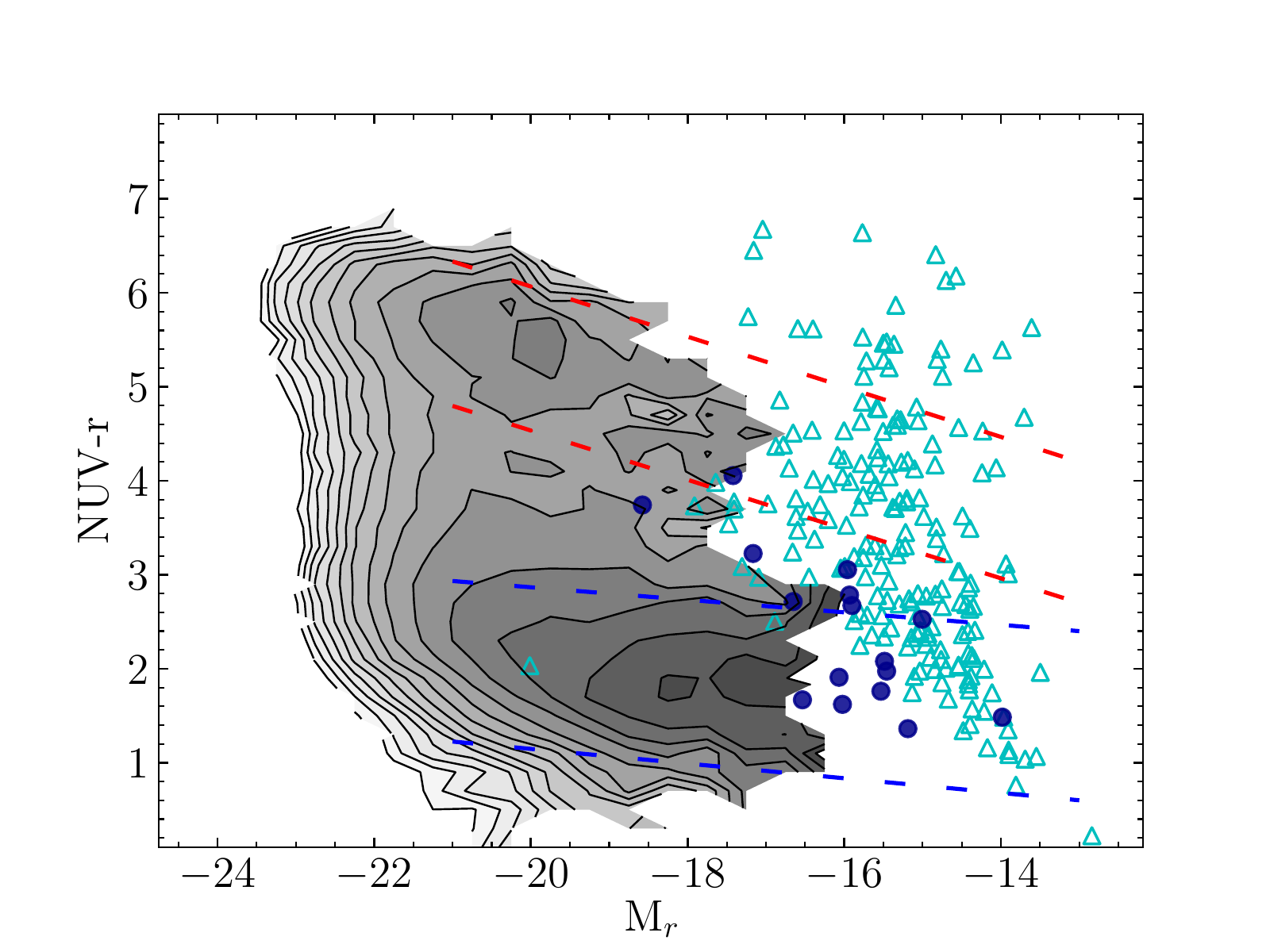}
\end{center}
\caption{A comparison of luminous galaxies and our UDGs. The isodensity contours represent the volume density of local galaxies in this color-magnitude space as derived by \cite{Colour_extn}. Our detections are shown as filled circles and the color limits from our non-detections are shown as upward pointing triangles and illustrate the allowed direction for those points to migrate. Our detections are mostly consistent with the field blue population, although pushing to fainter galaxies than examined in the previous study.}
\label{fig:cmd}
\end{figure}

In Figure \ref{fig:cmd} we compare the distribution of both our detections and limits for non-detections to the distribution measured for the local galaxy population by \cite{Colour_extn}, assuming that the candidate UDGs lie at the distance of the Coma cluster. \cite{smudges} provide evidence suggesting this is likely to be the correct distance for most, but not all, of the UDG candidates within this first released survey area. Errors in distance affect only the location of the candidate UDG along the abscissa and so do not significantly impact what we conclude next. We expect errors in our assigned distances to be mostly overestimates because placing the candidates beyond Coma would imply even larger physical sizes.

In Figure \ref{fig:cmd}, we outline with blue and red dashed lines possible extrapolations of the blue cloud and red sequence galaxy populations. The equations for the top and bottom red dashed lines are
\begin{align}
(NUV-r) &= -0.267M_r+0.733, \\
(NUV-r) &= -0.262M_r - 0.707,
\end{align}
and the equations for top and bottom blue dashed lines are
\begin{align}
(NUV-r) &= -0.067M_r+1.534, \\
(NUV-r) &= -0.078M_r-0.416,
\end{align}
respectively.

For the red sequence, we deliberately define an extreme possibility of a highly tilted red sequence, for which there is a hint in the \cite{Colour_extn} data and which provides the most optimistic possibility for detecting red sequence UDGs in {\sl GALEX} data. Even so, we have only one detection that might lie on the red sequence. It is evident from this comparison, that we should not expect to detect UDGs that are not forming stars in these {\sl GALEX} images. 

The bulk of the detections lie in an area of the diagram consistent with an extrapolation of the blue cloud, suggesting that at least these objects are forming stars. However, 135 of the 189 non-detections have lower limits on NUV$-r$ that place them redder than the blue cloud, demonstrating that the majority of the candidates are not star-forming, low mass galaxies. Selecting objects that are large in angular extent and have low central surface brightness in the optical ($\mu_{g,0} > 24$ mag arcsec$^2$) results in a sample that is dominated by non- or weakly-star forming galaxies in all environments. 

\subsection{Stacking Analysis}

To examine the nature of the non-detections in the extrapolated red and blue sequences of Figure \ref{fig:cmd}, we first average their flux limits and propagate uncertainties to determine if it is feasible to detect an object in the image stack of these candidates if they all have the maximum allowed flux. For the red sequence non-detections, we find that even a stack all of such sources will have
insufficient signal for a detection. For the blue sequence non-detections, excluding the one bright UDG candidate at $M_r \sim -20$, we find that we could detect the stacked object, provided the objects are not much fainter than the NUV limits. 

Our stack of the images of the UDG candidates whose NUV limits potentially place them in the blue cloud region of Figure \ref{fig:cmd} does result in a detected central object. The stack might include contaminating objects, which we have no way of excluding, and therefore our detection should be treated as a lower limit on NUV$-r$. Even so, our measurement (NUV$-r \sim 4.5$) places the mean object
on the red sequence rather than in the blue cloud, suggesting that the majority of 
the objects contributing to the stack must be on the red sequence. Because we now  exclude even most non-detections from the blue cloud, we conclude that the bulk of our UDG candidates are indeed quiescent. 

\begin{figure}[!htbp]
\begin{center}
\includegraphics[width = 0.48 \textwidth]{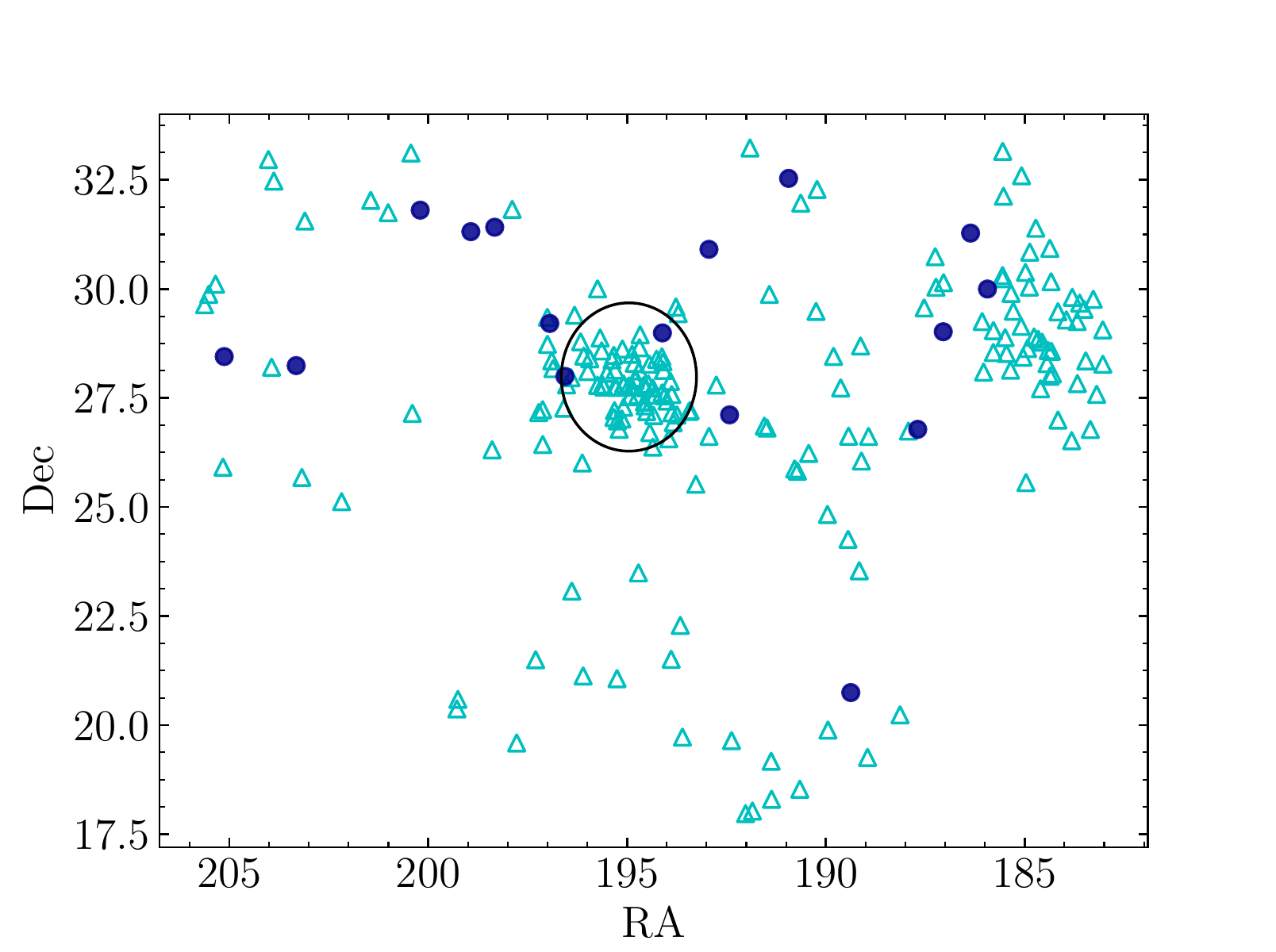}
\end{center}
\caption{The distribution on the sky of NUV detected (solid circles) and non-detected UDG candidates (open triangles). The Coma cluster is the central concentration of UDG candidates, with the large circle representing the virial radius. No NUV detections lie within the densest concentration and so NUV strong objects are manifestly underrepresented in this environment.}
\label{fig:skyplot}
\end{figure}

\subsection{Spatial Distribution}

In Figure \ref{fig:skyplot} we present the spatial distribution of the UDG candidates, coded by whether or not we detect NUV flux. The few detections that lie close to the Coma cluster in projection lie on the periphery. The candidate UDGs for which we were unable to make a measurement because they lacked either {\sl GALEX} imaging or $r-$band data are distributed randomly across the survey area and are not shown in the Figure. The UDG candidates outside of Coma are also predominantly non-detections. 

A concern in interpreting the distribution of non-detections is that a fraction of these sources were rejected due to possible contamination. As such, some could have significant NUV flux. Furthermore, it is possible that within regions of high galaxy density we are more likely to identify possible contaminating sources within the aperture. However, we visually rejected only 5 sources within the Coma virial radius. We rejected 16 outside of Coma. The vast majority of UDG candidates (110) within the Coma virial radius had no detectable flux. We conclude that the lack of candidates identified as NUV detections within the Coma cluster is not due to a bias caused by increased contamination within the measurement apertures.

A second potential concern is that the NUV survey depth is uneven across the field. However, the high density areas that proportionally lack NUV detections, the Coma cluster and the area toward the right of Figure \ref{fig:skyplot}, tend to have the deeper {\sl GALEX} imaging available than the low density areas. We conclude that survey depth variations are not responsible for the relative lack of NUV detections in dense environments. 

\begin{figure}[!htbp]
\begin{center}
\includegraphics[width = 0.48 \textwidth]{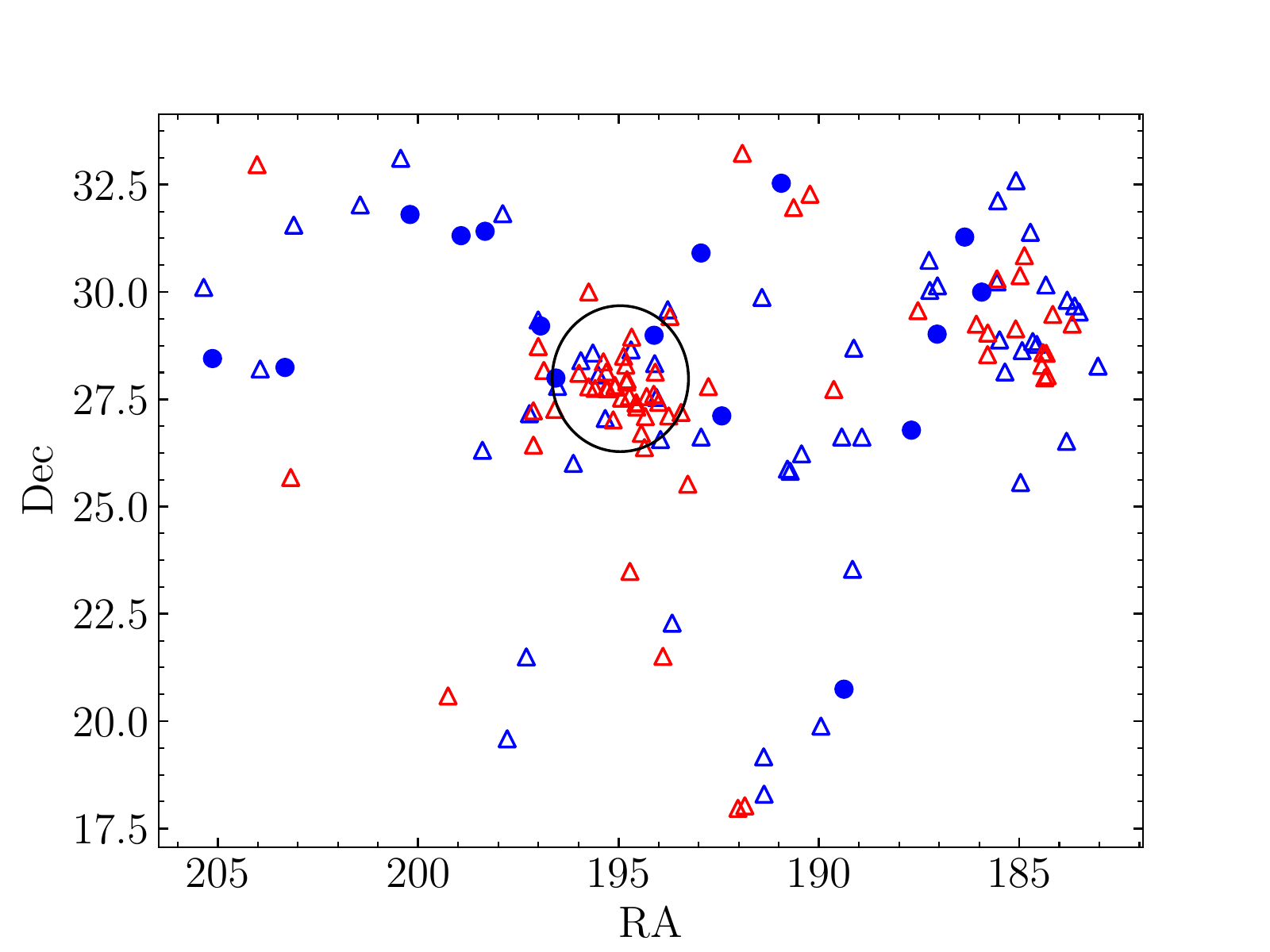}
\end{center}
\caption{The distribution on the sky of candidate UDGs that lie in either our extrapolated red sequence or blue cloud regions defined in Figure \ref{fig:cmd}. The colors of the symbols accurately reflect the candidate's association with either the red sequence or blue cloud. Filled circles represent NUV detections, open triangles represent non-detections. The large circle represents the Coma cluster virial radius.}
\label{fig:blue_red}
\end{figure}

In Figure \ref{fig:blue_red} we show the spatial distribution of a subset of UDG candidates, those that lie either within our extrapolated red sequence or blue cloud regions in Figure \ref{fig:cmd}. Because the majority of these points represents limits on the color, the blue cloud points could represent red UDG candidates, while the red sequence points are securely red. The Coma cluster is dominated by the presence of red sequence galaxies. 
This confirms the findings of \cite{remco} who, on the basis of optical colors, concluded that cluster UDGs are nearly all quiescent.  There is an apparent over representation of red UDG candidates also in the overdensity on the right of Figure \ref{fig:blue_red}, and a relative overabundance of blue points in the field. However, we know from our stacking analysis that the bulk of these are also quiescent galaxies.

A particularly interesting population of objects are those UDG candidates for which the NUV$-r$ color limit constrains them to be a quiescent field UDG candidate. These  are confidently identified as both quiescent and outside of a dense galaxy environment. Because we do not know their distance, we cannot confirm that they are indeed UDGs rather than dwarf elliptical/spheroidal galaxies. Nevertheless, there are $\sim$ 24 UDG candidates that lie  at least 3  virial radii in projected separation from Coma, beyond where cluster effects are observed and theorized to end \citep[cf.][]{gomez,zinger} and are confirmed as red for which redshifts would be particularly valuable. {\sl GALEX} screening of the larger samples expected from the SMUDGes survey will be an efficient way to identify quiescent field UDG candidates. In the entire sample, again assuming that these lie at the distance of the Coma cluster, 8 detections and 21 non-detections belong to UDGs that have $r_h>$ 4 kpc. Seven from the latter category lie well outside  both the Coma cluster and the galaxy concentration seen toward the right of Figure \ref{fig:blue_red}.

\subsection{Star Formation Rates}

Interpreting the NUV flux as originating from young stars, we proceed to estimate the stellar mass normalized star formation rate, or specific star formation rate (sSFR). These estimates provide an upper limit on the current sSFR because the NUV flux can also originate from somewhat older stars. We will return to this topic farther below when we discuss the FUV$-$NUV colors. Measurements of the sSFR were done for the local galaxy population by \cite{schiminovich}, and we follow their approach. We obtain the star formation rate (SFR) from the NUV flux using the relation
\begin{equation}
    \text{SFR (M$_{\odot}$ yr$^{-1}$)} = 1.0\times10^{-28} L_{\nu} \text{ (ergs s$^{-1}$ Hz$^{-1}$)}
\end{equation}
as adopted by \cite{Colour_extn}. To estimate the stellar mass, we first calculate the mass to light ratio (M/L) from the $g-z$ color using the relation 
\begin{equation}
    \log_{10}(M/L) = -0.171 + 0.322\times(g-z)
\end{equation}
from \cite{bell}.  The $g$ and $z$ magnitudes for the UDGs are given by \cite{smudges} and extinction corrected. The stellar mass is then estimated by taking the product of $M/L$ with the $z$ band luminosity. The \cite{bell} relation was not derived including UDG stellar populations, but the uncertainty in M/L is small in comparison to the range of SFRs. The sSFR is then simply given by the ratio of the SFR to the stellar mass.

We also calculated the sSFR using a more recent color stellar mass relation derived by \cite{roediger} to check the effect of using a different stellar mass estimator. Although, the new transformation noticeably moves the UDG candidates in the stellar mass - specific star formation rate space (the mean fractional change decreases the stellar mass by 46\%), the overall distribution remains almost the same and hence does not affect the qualitative results that we present next.

\begin{figure}[!htbp]
\begin{center}
\includegraphics[width = 0.48 \textwidth]{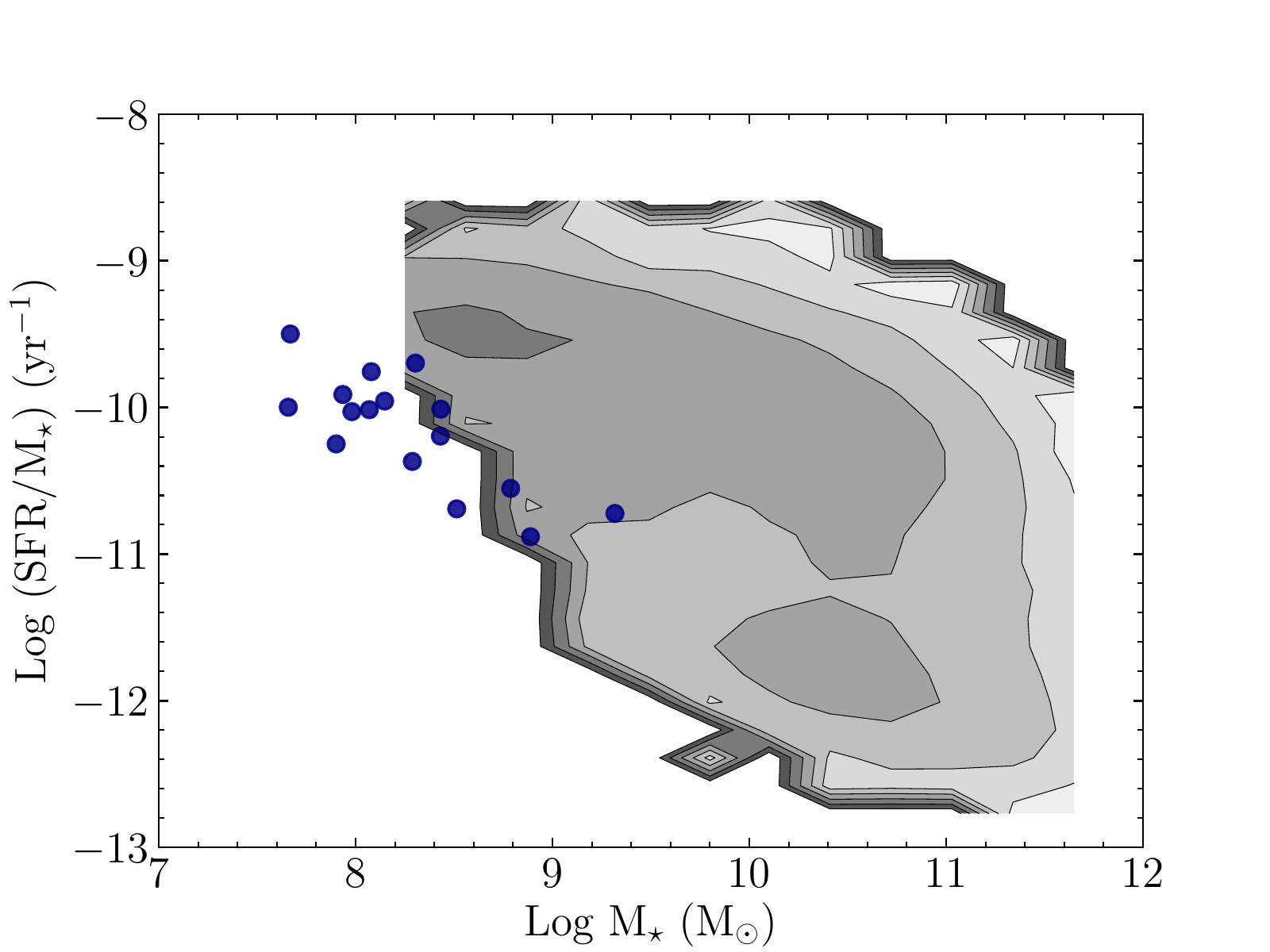}
\end{center}
\caption{The distribution of local galaxies and candidate UDGs in the stellar mass - specific star formation rate space. The isodensity contours represent the volume corrected distribution for local galaxies from \cite{schiminovich} and the blue dots represent our candidate UDG detections. We estimate the stellar masses using the relationships presented by \cite{bell} and the UDGs move by $\sim$ 0.33 upward and to the left if we use the relationships presented by \cite{roediger}.}
\label{fig:sfr}
\end{figure}

We compare the calculated sSFR's for the 16 detected UDG candidates to the distribution of values found for local galaxies by \cite{schiminovich} in Figure \ref{fig:sfr}. The distribution of UDGs outlines the upper boundary of the region in this diagram that is populated by the UDG candidates from the SMUDGes survey. At the bluest end, the detected UDGs with the highest sSFR's almost reach the star forming sequence, suggesting that at the lowest stellar masses we detect the galaxies can be hosting relatively high specific star formation rates and still satisfy our surface brightness selection criteria. However, at most $\sim$ 10 out of the 258 UDG candidates are in this category, demonstrating that the UDG selection criteria is not broadly detecting field star forming dwarfs. Although field UDGs, as defined by the SMUDGes criteria, are generally optically bluer than their cluster counterparts \citep{smudges},  they are not low mass galaxies on the star forming sequence. They might possibly tend to be lower stellar mass, and hence lower metallicity, systems. At the high stellar mass end, we are just able to detect in {\sl GALEX} images objects near the quiescent sequence. The remainder of our detections lie somewhere between the two extremes. The non-detections will fill in the lower left of this diagram, but significantly deeper NUV imaging is necessary to place those objects on this diagram. Distance errors, if the UDG candidates are actually closer than the assumed Coma cluster distance, will move objects to the left in the diagram but do not qualitatively affect our conclusions.

Returning to the question of whether the NUV flux originates from ongoing star formation, we examine the FUV$-$NUV color for the NUV-detected objects. All but two of the detections are red in FUV$-$NUV ($> 1$), suggesting that the sSFR's estimated using the NUV fluxes are overestimates of the current sSFR. 
Such red colors are consistent with those measured for LSB galaxies \citep{Boissier_2008, Wyder_2009}. Although the UDG colors appear to be somewhat redder on average than those published for the LSBs, the reddest among our systems have the largest uncertainties and, within 2$\sigma$, are consistent with the LSB colors.
Interpreting our measurements of sSFR as upper limits, places even more of the NUV-detected objects into the quiescent category. Only in two cases are the FUV$-$NUV colors consistent with ongoing star formation. This confirms our general result that the SMUDGes UDGs are almost exclusively non-star forming systems in all environments and is consistent with the previous determination that low surface brightness galaxies have lower star formation efficiencies than high surface brightness galaxies \citep{Wyder_2009}.

\section{Conclusions}

We present NUV measurements from {\sl GALEX} images of the NUV fluxes or flux limits for the sample of ultra-diffuse galaxies (UDGs) presented by the SMUDGes survey \citep{smudges}. Of the 258 UDG candidates that have the necessary {\sl GALEX} imaging and $r-$band photometry, we reject 53 as being strongly contaminated by a nearby source, measure a statistically significant NUV flux, and hence NUV$-r$ color, for 16, and place limits on the NUV flux for the remaining 189. 

For a subset of the non-detections the limits are sufficiently constraining that we conclude that they are quiescent galaxies. Furthermore, some of these lie well outside of any high galaxy density region, making those objects prime candidates for quiescent, field UDGs. It is important to measure the distances to these objects and determine whether these are indeed physically large galaxies. Models of UDG formation find it challenging to produce quiescent, large, field, low surface brightness galaxies. 

None of the NUV detected objects are in regions of high density, confirming conclusions reached using optical photometry that UDGs within clusters are quiescent. Furthermore, our analysis of the image stack of candidate UDGs whose photometric limits did not preclude them from being star forming objects yielded an NUV flux measurement that demonstrated that the majority of the objects in that stack are actually quiescent as well.  Lastly, only two of the NUV detected UDGs have FUV$-$NUV colors consistent with ongoing star formation. We conclude that the vast majority of the SMUDGes UDG candidates are quiescent, independent of environment. 

{\sl GALEX} archival imaging provides valuable ancillary information that can relatively easily help identify key sub populations from the large number of UDG candidates that ongoing surveys will identify. 

\section{Acknowledgments}

    We thank Remco van der Burg and Ananthan Karunakaran for helpful comments on an earlier draft. Some of the data presented in this paper were obtained from the Mikulski Archive for Space Telescopes (MAST). STScI is operated by the Association of Universities for Research in Astronomy, Inc., under NASA contract NAS5-26555. We gratefully acknowledge financial support for this work from the National Science Foundation (NSF AAG-171384)


\end{document}